\documentclass[prx,onecolumn,nofootinbib,citeautoscript,eqsecnum,10pt,longbibliography,notitlepage]{revtex4-2}
\usepackage{graphicx}
\usepackage{dcolumn}
\usepackage{bm}
\usepackage{color} 
\usepackage{amsmath}
\usepackage{tabularx,graphicx}
\usepackage{epstopdf}
\usepackage{latexsym}
\usepackage{amssymb}
\usepackage{amsmath}
\usepackage{color, colortbl}
\usepackage{psfrag}
\usepackage{bbm}
\usepackage{bm}
\usepackage{titlesec}
\usepackage{dsfont}
\usepackage{feynmp}
\usepackage{slashed}
\usepackage{multirow}
\usepackage[tight]{subfigure}

\usepackage[papersize={8.5in,11in}]{geometry}

\usepackage{color}
\definecolor{darkblue}{rgb}{0.,0.,0.4}
\definecolor{darkred}{rgb}{0.5,0.,0.}
\definecolor{BlueViolet}{RGB}{138,43,226}
\definecolor{SkyBlue}{RGB}{30,144,255}
\definecolor{DarkGreen}{RGB}{0,100,0}
\usepackage[pdftex,colorlinks=true,linkcolor=darkblue,citecolor=blue,urlcolor=darkred]{hyperref}

\usepackage{physics}

\newcommand{\bk}{\mathbf{k}}

\newcommand{\bq}{\mathbf{q}}

\newcommand{\bA}{\mathbf{A}}
\newcommand{\be}{\mathbf{e}}

\renewcommand{\epsilon}{\varepsilon}

\def \nn{\nonumber \\}

\begin{document}

\title{Raman response and shear viscosity in the non-Fermi liquid phase of Luttinger semimetals}

\author{Ipsita Mandal}
\affiliation{Institute of Nuclear Physics, Polish Academy of Sciences, 31-342 Krak\'{o}w, Poland}

\author{Hermann Freire}
\affiliation{Instituto de F{\'i}sica, Universidade Federal de Goi{\'a}s, 74.001-970,
Goi{\^a}nia-GO, Brazil}

\begin{abstract}
Luttinger semimetals represent materials with strong spin-orbit coupling, harbouring doubly-degenerate quadratic band touchings at the Brillouin zone center. In the presence of Coulomb interactions, such a system exhibits a non-Fermi liquid phase [dubbed as the Luttinger-Abrikosov-Beneslavskii (LAB) phase], at low temperatures and zero doping. However, a clear experimental evidence of this emergent state remains elusive to this date. Hence, we focus on extracting the Raman response as a complementary experimental signature. At frequencies much larger than the temperature, the Raman response exhibits a power-law behavior, which can be verified experimentally. On the other hand, at lower frequencies, the Raman response displays a quasi-elastic peak. We also compute the ratio of the shear viscosity and the entropy density, and the value obtained is a consequence of the hyperscaling violation that emerges in the LAB phase.
\end{abstract}

\maketitle

\tableofcontents

\section{Introduction}

Luttinger semimetals \cite{Luttinger} refer to a class of strongly spin-orbit coupled fermions with pseudospin-3/2 in three dimensions, whose normal state exhibits quadratic touching of Kramers-degenerate valence and conduction bands at the Brillouin zone center (i.e., the $\Gamma$-point). Due to the growing relevance of these strongly correlated compounds nowadays in the field of topological quantum materials, there has been a recent surge of theoretical works investigating various aspects of the physics of these systems \cite{moon-xu, rahul-sid, *ips-rahul,*ips-rahul-errata, lukas-herbut,igor16, ips-qbt-sc,ips_qbt_plasmons,ips_qbt_tunnel,Roy_PRB,Herbut-PRB,krempa,polini,Boettcher_2019,ipsfloquet, ips_hermann1,ips-hermann2}. Important examples of these compounds include HeTe \cite{Bernevig_2006,Kane_RMP,Zhang_RMP}, some pyrochlore iridates \cite{Witczak-Krempa,Kondo_2015}, grey-tin \cite{Groves}, and half-Heusler compounds \cite{Paglione,Taillefer_2013}.

Since the Luttinger semimetal harbors an isolated Fermi node at the Brillouin zone center, the electron-electron interactions are not effectively screened in these systems. As a consequence, a minimal model to describe these compounds must include strong (long-range) Coulomb interactions \cite{moon-xu,rahul-sid,*ips-rahul,*ips-rahul-errata}.
It is by now well-established that the interactions mediated by the Coulomb forces in Luttinger semimetals stabilize a new non-Fermi liquid (NFL) state -- the so-called Luttinger-Abrikosov-Beneslavskii (LAB) phase \cite{Abrikosov_Beneslavskii,Abrikosov}. The effective field theory for this phase was first studied by Abrikosov and Beneslavskii in the 1970s in a controlled approximation, by using a large-$N$ expansion \cite{Abrikosov_Beneslavskii,Abrikosov}. This NFL phase was later revisited and further reformulated using dimensional regularization and renormalization group (RG) techniques by Moon \emph{et al.} \cite{moon-xu}, with many interesting new predictions.
An important distinction of the LAB phase from other well-known NFLs arising for critical Fermi surfaces \cite{nayak,nayak1,lawler1,mross,Jiang,ips2,ips3,Shouvik1,Freire_RG_PDW,Lee-Dalid,shouvik2,Freire_Pepin_1,ips-uv-ir1,ips-uv-ir2,Freire_Pepin_2,ips-subir,ips-sc,ips-c2,Lee_2018,ips-fflo,ips-nfl-u1} is that the former represents an NFL fixed point at a Fermi node, rather than for a Fermi surface. From recent analytical works, we do have some other examples of nodal NFLs \cite{malcolm,ips-biref} as well.

Recently, we have calculated the optical conductivity, the dc conductivity, the thermal conductivity, and the thermoelectric response of the LAB phase, using the Kubo formula and the memory matrix approaches \cite{ips_hermann1, ips-hermann2}. Our theoretical results (when applicable) agree qualitatively with recent experimental data obtained for Luttinger semimetal compounds like the pyrochlore iridates [(Y$_{1-x}$Pr$_x$)$_2$Ir$_2$O$_7$] \cite{pramanik}. Therefore, as a further step to find other distinctive experimental consequences of the LAB phase, we calculate the Raman response in this paper. Nowadays, the Raman spectroscopy has become an indispensable experimental technique for studying strongly correlated materials (for a comprehensive review, see Ref. \cite{Devereaux_RMP}), since it is well-suited to yield valuable information about the non-trivial dynamics of the electronic excitations in these systems.

In view of the NFL nature of the LAB fixed point in $d=3$ for the Luttinger model, it is physically reasonable to assume that the effects of electron-phonon interaction and the electron-impurity coupling are negligible in these systems. Therefore, an effective hydrodynamic regime is expected to emerge at low temperatures. One key transport property that characterizes such a regime is the shear viscosity, which measures the dissipative effect related to the internal friction in the electron fluid. In a breakthrough paper in 2005 \cite{DTSon-PRL_2005}, Kovtun-Son-Starinets proposed that the ratio of the shear viscosity ($\eta$) and the entropy density ($s$) for hydrodynamic systems has a universal lower bound of the form $\eta/s \geq 1/(4\pi)$ ($\hbar$ and $ k_B$ are set to unity here), using the anti-de Sitter/conformal field theory correspondence. If $\eta/s$ turns out to be close to the lower bound of this inequality, then the system is classified as displaying ``minimal-viscosity'', and can be viewed as a measure of a strongly interacting system. Important systems that fit into the ``minimal-viscosity'' scenario include graphene at charge neutrality \cite{Fritz-PRB}, quark-gluon plasma \cite{DTSon-PRL}, and ultracold quantum gases tuned to the unitarity limit \cite{Cao}.

Not all strongly coupled quantum field theories conform to the lower bound of the Kovtun-Son-Starinets ratio though. Recently, Patel \textit{et al.} \cite{patel2} has shown that a diverging $\eta/s$ can appear at low temperatures, in the context of an NFL appearing at the Ising-nematic quantum critical point \cite{Lee-Dalid,ips-uv-ir1,ips-uv-ir2}. This behavior can be traced to the violation of the hyperscaling property of the corresponding field theory model \cite{ips-subir}. Technically speaking, the hyperscaling is a property in which the entropy scales as if the system were defined effectively in $(d-\theta)$ dimensions, where $\theta$ is the so-called hyperscaling violation exponent. Indeed, there have been many studies of quantum critical points, which display varying degrees of hyperscaling violation \cite{patel2,Dong_2012,Karch_2014}. There are essentially two ways through which this can happen: (1) due to a violation of the naive scaling of the conductivity $\sigma(\omega=0,T)$ (where $T$ denotes the temperature), which comes together with the naive scaling violation of $s$;
(2) due to a mismatch between the scaling in $\omega$ of $\sigma(\omega,T=0)$ and the $T$-dependence of $\sigma(\omega=0,T)$. The second scenario usually appears when there is a dangerously irrelevant deformation in the corresponding low-energy effective field theory, breaking the charge-conjugation symmetry of the (otherwise charge-conjugation symmetric) infrared fixed point \cite{Blaise,Blaise2}.

Recently, we have shown that the LAB phase violates the hyperscaling property at low temperatures \cite{ips_hermann1}, by calculating the optical conductivity of the model.
In this paper, we continue to investigate this NFL phase further by calculating the $\eta / s $ ratio, using the memory matrix method. From our calculations, we find that this ratio scales in $d=3$ as $\eta/s \sim T^{\lambda-49/38}$, where $0<\lambda<1$. Therefore, we find that $\eta/s$ diverges in three dimensions as $T\rightarrow 0$, instead of saturating to a finite value.

The paper is organized as follows. In Sec.~\ref{model}, we introduce the model for the Luttinger semimetal, augmented by long-ranged Coulomb interactions. We also describe the low-energy NFL fixed point characterizing the LAB phase, obtained from a dimensional regularization scheme in $d=4-\varepsilon$ spatial dimensions. In Sec.~\ref{secraman}, we compute the Raman response at $T=0$ using the Kubo formalism, and also at $T>0$ using the memory matrix approach.
We then calculate the free energy, specific heat, and the entropy density in Sec.~\ref{Free_energy}.
Finally, we derive the expressions for the optical ($T=0$) and dc ($T>0$) viscosities in Sec.~\ref{shear_visc}. This allows us to work out the $\eta/s$ ratio in Sec.~\ref{secetabys}.
We end with some concluding remarks in Sec.~\ref{secsum}.

\section{Model}
\label{model}

We consider the isotropic Luttinger Hamiltonian \cite{Luttinger} captured by
\begin{equation}\label{bare2}
 \mathcal{H}_0 = \frac{|\mathbf{k}|^2}{2\, m'}-\frac{\frac{5\, |\mathbf{k}|^2 } {4}
 -(\mathbf{k}\cdot\mathbf{J})^2}{2\,m}\,,
\end{equation}
where $\mathbf{J}$ is the vector angular momentum operator for pseudospin-$3/2$ states. This effective Hamiltonian emerges from the three-dimensional band-structure of certain spin-orbit coupled systems, and harbors quadratic band crossings at the Brillouin zone center. The energy eigenvalues are
\begin{align}
\epsilon_\pm (\mathbf k) = \frac{|\mathbf{k}|^2}{2 \,m'} \pm \frac{|\mathbf{k}|^2}{2 \,m}\,,
\end{align}
where the ``+'' and ``-'' signs refer to the conduction and valence bands, respectively. Each of these bands is doubly degenerate.

The model can also be expressed in an equivalent form \cite{moon-xu,rahul-sid,*ips-rahul,*ips-rahul-errata,ips_hermann1}
\begin{equation}
\label{bare}
 \mathcal{H}_0 = \sum_{a=1}^5 d_a(\mathbf{k}) \,  \Gamma_a   
 + \frac{ |\mathbf{k}|^2}{2\,m'} \,,
\quad d_a  (\mathbf{k})   = \frac{\tilde d_a(\mathbf{k})   }{2\, m}\,,
 \end{equation}
using $m$ and $m'$ as the mass parameters that define the quadratic bands.
The $\Gamma_a$ matrices provide 
a rank-four irreducible representation of the Euclidean Clifford algebra, obeying the anticommutation relation $\{\Gamma_a,\Gamma_b\} = 2\, \delta_{ab}$. Furthermore, $\tilde d_a(\mathbf k)$'s represent the angular momentum $l=2$ spherical harmonics \cite{Murakami}, which have the explicit forms as follows:
\begin{align}
\label{ddef}
&\tilde d_1(\mathbf{k}) = \sqrt{3}\, k_y \,k_z\,,
\quad \tilde d_2(\mathbf{k}) =  \sqrt{3}\, k_x\, k_z\, ,\quad
 \tilde d_3(\mathbf{k}) =  \sqrt{3} \,k_x\, k_y\, ,\quad
 \nn & 
 \tilde d_4(\mathbf{k}) =\frac{\sqrt{3}  \,  (k_x^2 - k_y^2) }{2}\,, \quad
 \tilde d_5(\mathbf{k}) = \frac{2\, k_z^2 - k_x^2 - k_y^2}{2} \,.
\end{align}
We will sometimes use the notation $\mathbf{d}_{\mathbf{k}}$ instead of $\mathbf{d}(\mathbf{k})$ to avoid cluttering in long equations.

The term $\frac{ |\mathbf{k}|^2} {2\,m'}$ multiplies an identity matrix, causing the band masses of the conduction and valence bands to become unequal.
For simplicity, we will consider the case with $m'$ set to infinity (i.e. with equal band-masses) for all calculations \footnote{
At the LAB fixed point, the term $|\mathbf k|^2/(2\,m')$ in the Hamiltonian is irrelevant in the RG sense, as shown explicitly
in Ref.~\cite{rahul-sid}. Hence, the extremely involved calculation with a finite $m'$ will not give any result with significantly different physics.}, except for the Raman response.

Adding the Coulomb interactions via a non-dynamical scalar boson field $\varphi$, the action of this system\footnote{We have used here $N_f$ fermionic flavors, which allows one to check the loop-calculations using an alternative approach, namely the large-$N_f$ limit \cite{Abrikosov,moon-xu}. The physical scenario is given of course by $N_f=1$.} can be straightforwardly written as
\begin{align}
S_0 =&  \int d\tau \,d^3{\mathbf x}
\left[ \sum_{i=1}^{N_f} 
\psi_i^{\dag}
 \left(  \partial_{\tau} + \mathcal{H}_0 + \mathrm{i}\, e 
 \,\varphi \right) \psi_i +\frac{c}{2} \big( \nabla \varphi  \big)^2  \right ].
\end{align}
Here, $\psi_i$ denotes the fermionic spinor with flavor index $i$, and $c$ is a constant equal to $1/(4 \,\pi)$.

The bare Green's function for each fermionic flavor is given by
\begin{align}
G_0(k_0, \mathbf{k}) =  
\frac{ \mathrm{i}\, k_0- \frac{ \mathbf k^2}{2\,m'}  + \mathbf{d}(\mathbf{k}) \cdot{\mathbf{\Gamma}}}
{-\left ( \mathrm{i}\, k_0- \frac{\mathbf k ^2}{2\, m'}  \right )^2 +|\mathbf{d}(\mathbf{k})|^2}\,,
\label{baregf}
\end{align}
where $|\mathbf{d}(\mathbf{k})|^2 = \frac{ \mathbf k^4} {4\,m^2}$.

\begin{figure}
\includegraphics[width=0.35\textwidth]{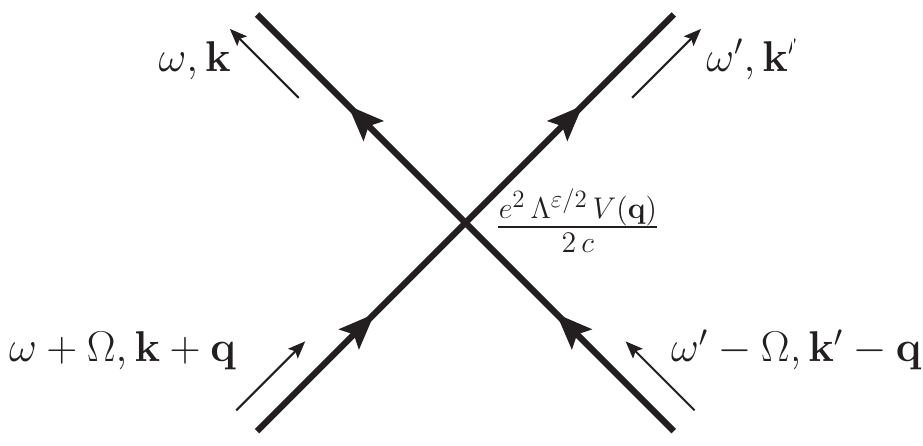}
\caption{The Coulomb interactions give rise to a four-fermion vertex.
\label{fig:vert}}
\end{figure}

We now integrate out $\varphi$ to express the Coulomb interaction as an effective four-fermion interaction vertex, resulting in the final action 
\begin{align}
S=& \sum_{i=1}^{N_f} 
\int \frac{d\tau \,d^d {\mathbf k} } {(2\,\pi)^d}\, 
{ \psi}_i^{\dag}(\tau,\mathbf k)
\left(\partial_{\tau} + \mathcal{H}_0 \right)
 { \psi}_i  (\tau, \mathbf k) \nonumber\\
& + \frac{e^2\,\Lambda^{\varepsilon}}{2 \,c }
 \sum_{i,j=1}^{N_f}
\int \frac{d\tau\,
d^d {\mathbf k}\, d^d {\mathbf k'}\,d^d {\mathbf q} }
{(2\pi)^{3d}}\, 
V(\mathbf q)\,{\psi}^{\dag}_{i} (\tau,{\mathbf k}+\mathbf q)
\,{\psi}_{i}(\tau,{\mathbf k})\,
 {\psi}^{\dag}_{j}( \tau ,{\mathbf k}'-\mathbf q)\, 
\psi_j (\tau,{\mathbf k}')   \,,
\label{action}
\end{align}
in $d$ spatial dimensions, where
where $V(\mathbf q) = 1/\mathbf q^2$.
In the momentum space, the Coulomb interaction vertex is given by
$\frac{e^2\, \Lambda^{\varepsilon}}{2 \,c }V(|\mathbf q|)$, as shown in Fig.~\ref{fig:vert}.
We have also scaled $\frac{e^2}{c}$ by using the floating mass scale $\Lambda$ (of the renormalization group flow) to make it dimensionless for $d=4-\varepsilon $ spatial dimensions.
\footnote{Here, we fix the tree-level scaling dimension by setting the scaling dimension $[k]$ as $1$. Using this, the various tree-level scaling dimensions are given by: $[\tau]=-z =-2$ (where $z$ is the dynamical critical exponent), $[1/m] = z-2$, and $[e^2] = 2\,z-d $ (before using the scaling factor $\Lambda^\varepsilon$).}
 
This action has an NFL phase, which can be accessed in a controlled way using dimensional
regularization, by considering the theory in $d$ spatial dimensions. The renormalization flow equation for the coupling constant $e$ gives a stable interacting fixed point \cite{moon-xu,Abrikosov}, with the value
\begin{align}
e^*= \sqrt{ \frac{ 60\,\pi^2\,c\,\varepsilon} 
{m\,\left(4+15\,N_f\right)} } \,.
\label{eqe}
\end{align}
This is an NFL with no long-lived quasiparticles at low energies, describing the LAB phase, with the dynamical critical exponent $z$ at the fixed point given by $z^*=2-\frac{4\,\varepsilon} {4 + 15\, N_f }$.

For performing the integrals without a finite UV cutoff, we will employ the scheme developed by Moon {\it et al.} \cite{moon-xu}, where the radial momentum integrals are performed with respect to a $d=4-\varepsilon$ dimensional measure $\int 
\frac{ |\mathbf k|^{3-\varepsilon} d|\mathbf k| }
{(2\,\pi)^{4-\varepsilon}}$, but the $\Gamma$ matrix structure of $d=3$ is retained. Although the angular integrals are performed only over the three-dimensional sphere parameterized by the polar and azimuthal angles $(\theta, \,\phi)$, the overall angular integral of an isotropic function $\int_{\hat{\Omega}}\cdot 1$ is taken to be $2 \,\pi^2$ (which is appropriate for the total solid angle in $d=4$), and the angular integrals are normalized accordingly. Therefore, the angular integrations are performed with respect to the following measure: 
\begin{equation}
\label{moonmeasure}
\int dS\, (\ldots) \equiv \frac{\pi}{2} \int_0^{\pi} d \theta \int_0^{2\,\pi} 
d \phi\, \sin \theta \, (\ldots)\,,
\end{equation}
where the $\pi/2$ is inserted for the sake of normalization.
To perform the full loop integrals, we will use the relations shown in Appendix \ref{angular}. For evaluating various traces involving the gamma matrices, we will use the identity for the trace of the product of four gamma matrices, i.e.,
\begin{align}
\Tr\left[\Gamma_a\,\Gamma_{a'}\,\Gamma_b\,\Gamma_{b'}\right]
=4\left( \delta_{aa'}\, \delta_{bb'}- \delta_{ab}\, \delta_{a'b'}
+ \delta_{ab'}\, \delta_{a'b}\right),
\end{align}
and also the standard identity for the trace of six gamma matrices [cf. Eq.~\eqref{eq6trace}]. Lastly, in the rest of the paper, we will set $N_f=1$.

\section{Raman response}
\label{secraman}

\begin{figure*}[t]
	\centering
	\subfigure[]{\includegraphics[width=0.15\textwidth]{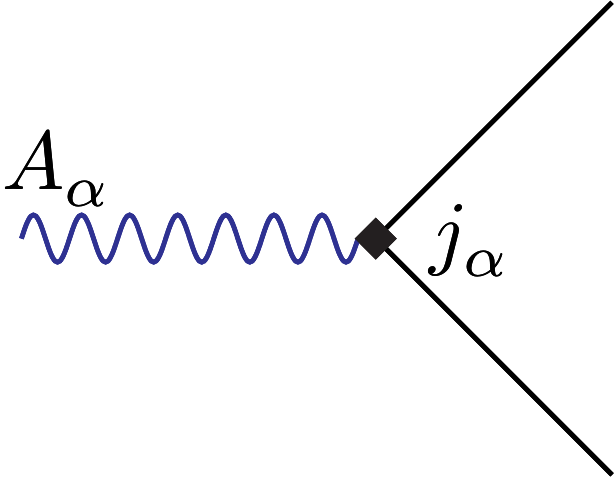} \label{figtree1}}\hspace{3 cm}
		\subfigure[]{\includegraphics[width=0.12 \textwidth]{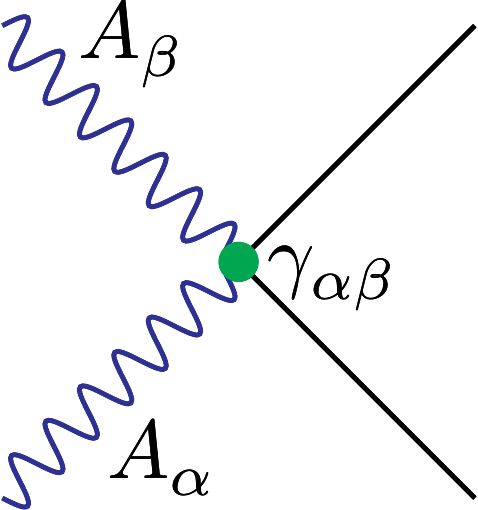} \label{figtree2}}
	\caption{\label{figvert}Electron-photon coupling vertices for the Raman response. The blue curly lines represent the incident and scattered photons. (a) The first type of vertex represents coupling the electron’s current to a single photon, and is denoted with a black square. (b) The second type of vertex represents coupling the electron’s charge to two photons, and is denoted by a green dot.}
\end{figure*}

In this section, we calculate the Raman response of the LAB phase. 
Raman response involves coupling the electrons to an electromagnetic field representing the
incoming and outgoing photons. Hence,
we introduce a gauge coupling in the model via the Peierls substitution
$\bk \rightarrow \bk + e\,\bA/c$, where $\bA$ is the vector potential (this is also equivalent to introducing a scalar field $\phi$, such that $\bA=\nabla \phi$). As a result, the vector potential dependent terms are given by:
\begin{align}
  H_{\mathbf{A}} & 
= \frac{e}{c} \,A_\alpha\, j_\alpha(\bk) 
 + \frac{e^2 }{2\, {c^2}}\,A_\alpha \,A_\beta\, \gamma_{\alpha\beta}(\bk)\,,
\end{align}
where $ j_\alpha =\frac{k_\alpha} {m'} 
+ \partial_{k^\alpha} d_a(\bk)\,  \Gamma_a $
represents the components of the current operator, and
\begin{align}
& \gamma_{xx} =\frac{\Gamma_0}{m'} +\frac{\sqrt{3}\,\Gamma_4-\Gamma_5}{2\,m}\,,\quad
\gamma_{yy} = \frac{\Gamma_0}{m'} - \frac{\sqrt{3}\, \Gamma_4+\Gamma_5}{2\,m}\,,\quad 
\gamma_{zz} = \frac{\Gamma_0}{m'} +\frac{\Gamma_5}{m}\,,\nn
& {\gamma_{xy}= \gamma_{yx}= \frac{\sqrt{3}\, \Gamma_3}{m}\,,\quad
\gamma_{yz}= \gamma_{zy}= \frac{\sqrt{3}\, \Gamma_1}{m}\,,\quad
\gamma_{xz}= \gamma_{zx}= \frac{\sqrt{3} \,\Gamma_2}{m}} \,.
\end{align}
There are two types of electron-photon couplings, as illustrated in Fig.~\ref{figvert}.
We note that in evaluating the loop diagrams for Raman response, the two vertices can only be either $(\Gamma_0,\Gamma_0)$, or $(\Gamma_a,\Gamma_b)$ for $a,b=1\dots 5$, as the cross-terms vanish identically.

The vector potential can be quantized as:
\begin{align}
    \bA ({\bq}) = \sqrt{\frac{h \,c^2} {\omega_{\bq}\,\mathcal V}} 
\left(\hat{\be}_\bq \,a_{-\bq} + \hat{\be}_{\bq}^* \,a_{\bq}^\dagger \right),
\end{align}
where $\mathcal V$ the volume, and
$a_{\bq}^\dagger$ ($a_{\bq} $) is the creation (annihilation) operator of the photons with energy
$\hbar \,\omega_{\bq} = \hbar\,c\,|\mathbf q|$ having a polarization direction denoted by $\hat{\be}_\bq$.
Within the Born approximation, the Raman scattering cross-section is given by \cite{Devereaux_RMP}:
\begin{align}
    \frac{\partial^2\sigma}
 {\partial \Omega \,\partial \omega_i} \propto \sum_{F,I} \frac{\exp(-\beta E_I)}{Z}\, |M_{FI}|^2\,\delta(E_F+\omega_s-E_I-\omega_i)\,,
\end{align}
where $(I,F)$ denote initial and final states of the Luttinger semimetal,
$Z$ is the partition function, and $ M_{FI} =\langle F \,|\,M\,|\,I \rangle $ (with $M $ being the effective light-scattering operator). The summation represents a thermodynamic average over all
possible initial and final states of the system having energies $E_I$ and $E_F$, respectively,
with the momentum vectors in the solid angle element $d\Omega$. 
Furthermore, $\omega =\omega_i-\omega_s $ is the net frequency, and $\mathbf q$ is the net momentum transferred by the photons.

Denoting the photon polarization vectors for incident and scattered light by $\hat{\be}_i$ and $\hat{\be}_s$, respectively, the Raman scattering amplitude is given by
\begin{equation}
    M_{FI}(t) = \bra{\hat{\be}_s,F}_t U(t,0) \ket{\hat{\be}_i,I}_0 \,,
\end{equation}
where in the interaction picture, we have
\begin{equation}
 U(t,0) = \mathcal{T}\, e^{-{\rm i} \int_0^t dt \,H_\bA(t)}\,,\quad
 H_\bA(t) = e^{{\rm i}\,\mathcal{H}_0 \,t}\,H_{\bA} \,e^{-{\rm i} \,\mathcal{H}_0\,t}\,.
\end{equation}

Let us define the operators
\begin{align}
{\rho}_0=\psi^{\dagger}\,\psi\,,\quad
{\rho}_a=\psi^{\dagger}\,\Gamma_a\,\psi\,,
\end{align}
whose two-point correlators will contribute to the Raman response $|M_{FI}|^2$ (see below).
Considering Raman scattering in the visible range, the zero momentum limit for the response is a good approximation \cite{Devereaux_RMP}. Hence, we will compute the correlators $\left \langle \rho_0\,\rho_0 \right \rangle (\mathrm{i}\,\omega)$ and $\left \langle \rho_a\,\rho_b \right \rangle (\mathrm{i}\,\omega)$ upto two-loop order.

Feynman diagrams for computing $ |M_{FI}|^2$ involve vertices of two types, as depicted in Fig.~\ref{figvert}. However, only diagrams consisting solely of green vertices involve non-resonant
scatterings, while the others give rise to resonant and mixed scatterings which can be neglected in the low-energy limit \cite{Devereaux_RMP}. Hence, the leading-order Feynman diagrams for $|M_{FI}|^2$ in the non-resonant scattering limit are only considered here.

\subsection{Raman response at $T=0$}

We now proceed with a perturbative calculation of the Raman response at $T=0$.

\subsubsection{One-loop contribution}

\begin{figure*}[]
	\centering
\includegraphics[width=0.25\textwidth]{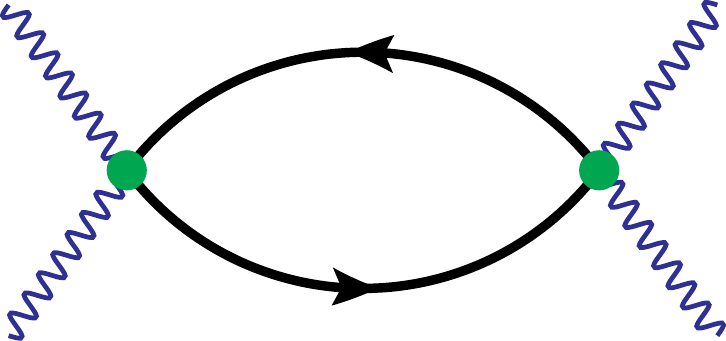}
\caption{Feynman diagram for the contribution to the Raman response at one-loop order.
\label{fig1loopraman}}
\end{figure*}

At one-loop order, from Fig.~\ref{fig1loopraman}, we obtain
\begin{align}
& \left \langle \rho_0\,\rho_0 \right \rangle_\text{1loop}(\mathrm{i}\,\omega)
= -  
 \int \frac{dk_0} {2\,\pi} \int \frac{d^d {\mathbf k}}{(2\,\pi)^{d}}
\text{Tr} \left [ G_0(k+q) \, G_0(k)\right ]\nn
&=  -  \int \frac{dk_0}{2\,\pi} \int \frac{d^d {\mathbf k}}{(2\,\pi)^{d}} 
\text{Tr} \left [ \frac{ \mathrm{i}\, k_0+\mathrm{i}\,\omega
- \frac{ \mathbf k^2}{2\,m'}  + \mathbf{d}(\mathbf{k}) \cdot{\mathbf{\Gamma}}}
{-\left ( \mathrm{i}\, k_0 + \mathrm{i}\,\omega- \frac{\mathbf k ^2}{2\, m'}  \right )^2 +|\mathbf{d}(\mathbf{k})|^2}
 \, \frac{ \mathrm{i}\, k_0- \frac{ \mathbf k^2}{2\,m'}  + \mathbf{d}(\mathbf{k}) \cdot{\mathbf{\Gamma}}}
{-\left ( \mathrm{i}\, k_0- \frac{\mathbf k ^2}{2\, m'}  \right )^2 +|\mathbf{d}(\mathbf{k})|^2}
\right ]\nn
&= - 4 
  \int \frac{dk_0}{2\,\pi} \int \frac{d^d {\mathbf k}}{(2\,\pi)^{d}}
\frac{  \left( \mathrm{i}\, k_0+\mathrm{i}\,\omega
- \frac{ \mathbf k^2}{2\,m'}  \right)
\left( \mathrm{i}\, k_0- \frac{ \mathbf k^2}{2\,m'}  \right)
+|\mathbf{d}(\mathbf{k})|^2 
}
{ \left [ -\left ( \mathrm{i}\, k_0 +\mathrm{i}\,\omega - \frac{ \mathbf k^2}{2\,m'}
\right )^2 +|\mathbf{d}(\mathbf{k})|^2
\right ] 
\left [ -\left ( \mathrm{i}\, k_0 - \frac{ \mathbf k^2}{2\,m'}  \right )^2 
+|\mathbf{d}(\mathbf{k})|^2 \right] } = 0 \text{ for } m'>m\,,
\label{eq:rr_1Loop0}
\end{align}
where we define henceforth $q=(\omega,\,\mathbf 0)$.

As for $a, b\neq 0$, we have:
\begin{align}
& \left \langle \rho_a\,\rho_b \right \rangle_\text{1loop}(\mathrm{i}\,\omega)
= -  
  \int \frac{dk_0}{2\,\pi} \int \frac{d^d {\mathbf k}}{(2\,\pi)^{d}}
\text{Tr} \left [  \Gamma_a\,
G_0(k+q) \, \Gamma_b \, G_0(k)\right ]\nn
&=  -  \int \frac{dk_0}{2\,\pi} \int \frac{d^d {\mathbf k}}{(2\,\pi)^{d}}
\text{Tr} \left [ \Gamma_a\,
\frac{ \mathrm{i}\, k_0+\mathrm{i}\,\omega
- \frac{ \mathbf k^2}{2\,m'}  + \mathbf{d}(\mathbf{k}) \cdot{\mathbf{\Gamma}}}
{-\left ( \mathrm{i}\, k_0 + \mathrm{i}\,\omega- \frac{\mathbf k ^2}{2\, m'}  \right )^2 +|\mathbf{d}(\mathbf{k})|^2}
 \, \Gamma_b\,
\frac{ \mathrm{i}\, k_0- \frac{ \mathbf k^2}{2\,m'}  + \mathbf{d}(\mathbf{k}) \cdot{\mathbf{\Gamma}}}
{-\left ( \mathrm{i}\, k_0- \frac{\mathbf k ^2}{2\, m'}  \right )^2 +|\mathbf{d}(\mathbf{k})|^2}
\right ]\nn
&= - 4 
  \int \frac{dk_0}{2\,\pi} \int \frac{d^d {\mathbf k}}{(2\,\pi)^{d}}
\frac{ 
\delta_{ab}
\left[ \left( \mathrm{i}\, k_0+\mathrm{i}\,\omega
- \frac{ \mathbf k^2}{2\,m'}  \right)
\left( \mathrm{i}\, k_0
- \frac{ \mathbf k^2}{2\,m'}  \right)
-|\mathbf{d}(\mathbf{k})|^2 \right]
+ 2\,d_{a}(\mathbf k)\, d_{b}(\mathbf k)
}
{ \left [ -\left ( \mathrm{i}\, k_0 +\mathrm{i}\,\omega - \frac{ \mathbf k^2}{2\,m'}
\right )^2 +|\mathbf{d}(\mathbf{k})|^2
\right ] 
\left [ -\left ( \mathrm{i}\, k_0 - \frac{ \mathbf k^2}{2\,m'}  \right )^2 
+|\mathbf{d}(\mathbf{k})|^2 \right] }
\nn & = -\frac{   m^{2-\frac{\varepsilon }{2}}\, |\omega|^{1-\frac{\varepsilon }{2}}
\,\delta_{ab} }
{10\, \pi }\,.
\label{eq:rr_1Loop}
\end{align}
Consequently, at zeroth order, the Raman response is proportional to $\omega^{1-\frac{\varepsilon }{2}}$ for $\omega\gg T$.

\subsubsection{Two-loop contributions}

\begin{figure*}[]
	\centering
	\subfigure[]{\includegraphics[width=0.25\textwidth]{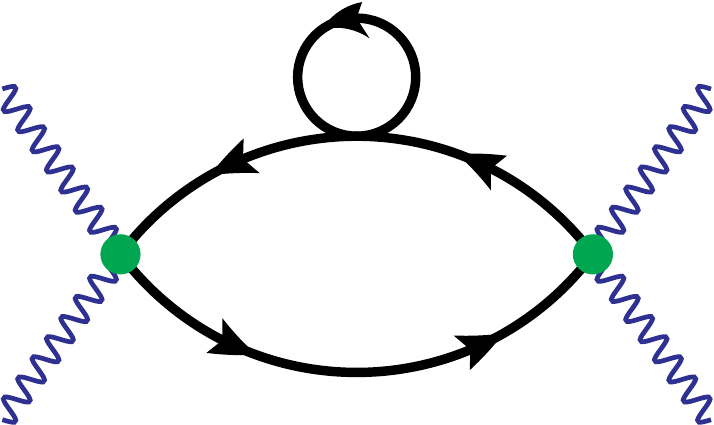} \label{figr2}}\hspace{1 cm}
	\subfigure[]{\includegraphics[width=0.25\textwidth]{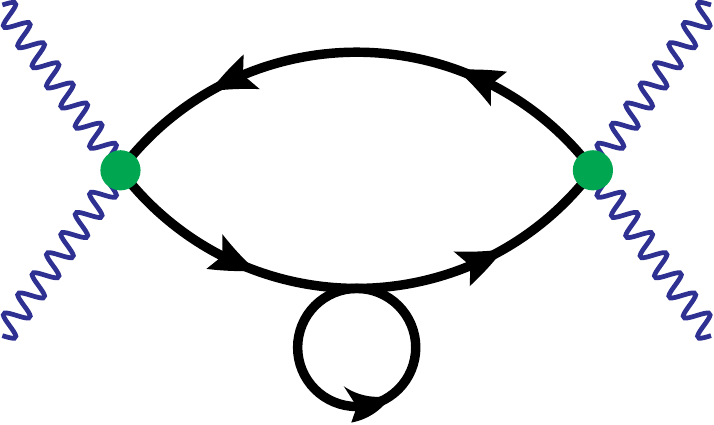} \label{figr3}}\hspace{1 cm}
	\subfigure[]{\includegraphics[width=0.25\textwidth]{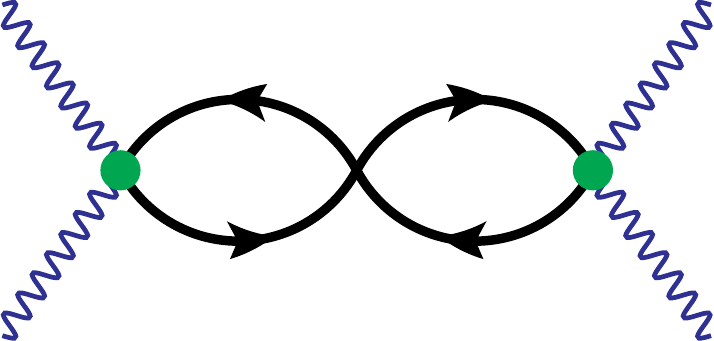} \label{figr4}}
	\caption{\label{fig2loopraman}Feynman diagrams for the contributions to the Raman response at two-loop order, with (a) and (b) representing the self-energy corrections, and (c) depicting the vertex correction.}
\end{figure*}


The first two-loop-order correction involves inserting the one-loop fermion self-energy ($\Sigma_1$) corrections into the correlator [cf. Fig.~\ref{figr2} and \ref{figr3}], which takes the form:
\begin{align}
 \left \langle \rho_a\,\rho_b \right \rangle_\text{2loop}^{(1)}(\mathrm{i}\,\omega)
&= -   \int \frac{dk_0}{2\,\pi} 
\int \frac{d^d {\mathbf k}}{(2\,\pi)^{d}}
\text{Tr} \left [ \Gamma_a\,G_0(k+q) \,\Sigma_1(\mathbf k)
\,G_0(k+q)\, \Gamma_b\, \,G_0(k)
\right ]\nn
&\quad -   \int \frac{dk_0}{2\,\pi} 
\int \frac{d^d {\mathbf k}}{(2\,\pi)^{d}}
\text{Tr} \left [ \Gamma_a\,G_0(k+q) \,\Gamma_b\,
G_0(k)\,\Sigma_1(\mathbf k)  \,G_0(k)
\right ] ,
\end{align}
where $\Sigma_1 (\boldsymbol \ell) 
= -\frac{ m\,e^2  } 
{15\,\pi^2\,c} \left( \frac{\Lambda}{|\boldsymbol \ell|} \right)^\varepsilon  
\frac{ \mathbf{d}(\boldsymbol \ell) \cdot \mathbf{\Gamma} }
{\varepsilon}$ (see Ref.~\cite{rahul-sid,*ips-rahul,*ips-rahul-errata}).
Evaluating the integrals for $a,b>0$ (for more details, see Appendix \ref{AppendixB}), we obtain:
\begin{align}
 \left \langle \rho_a\,\rho_b \right \rangle_\text{2loop}^{(1)}(\mathrm{i}\,\omega)
&=   
\frac{e^2  \,m^{3-\frac{\varepsilon }{2}}\, | \omega | ^{1-\frac{\epsilon }{2}} 
}
{ 75\, \pi ^3 \,c}
 \left[\frac{1}{\varepsilon }-\frac{1}{2} 
\ln \left(\frac{m \,| \omega | }{\Lambda^2}\right)\right]\,\delta_{ab}\,.
\end{align}

The second two-loop-order correction involves the vertex correction [cf. Fig.~\ref{figr4}], which can be expressed as:
\begin{align}
  \left \langle \rho_a\,\rho_b \right \rangle_\text{2loop}^{(2)}(\mathrm{i}\,\omega)
& =  \frac{e^2\,\Lambda^{\varepsilon}}{ c}
  \int \frac{dk_0\,d\ell_0}  {(2\,\pi)^2 }
\int \frac{d^d {\mathbf k}\, d^d {\boldsymbol{\ell}}}{(2\,\pi)^{2d}}
\text{Tr} \left [  \Gamma_a\,
G_0(k+q) \,\frac{1} {\boldsymbol{\ell}^2}\, G_0(k+q+\ell)\,
 \Gamma_b\, G_0(k+\ell)\,G_0(k) \right ]\nn
& = \frac{e^2\,\Lambda^{\varepsilon}} {c}
\int \frac{dk_0\,d\ell_0}  {(2\,\pi)^2 }
\int \frac{d^d {\mathbf k}\, d^d {\boldsymbol{\ell}}}{(2\,\pi)^{2d}}
\text{Tr} \left [  \Gamma_a\,
G_0(k_0+\omega,\mathbf k) \,\frac{1} {\boldsymbol{\ell}^2}\, 
G_0( \ell_0 + \omega,\mathbf k+\boldsymbol \ell)\,
 \Gamma_b \,G_0(\ell_0,\mathbf k +\boldsymbol \ell)
\,G_0(k_0,\mathbf k)
\right ],
\end{align}
with $q=(\omega,\mathbf 0)$ and $\ell =(\ell_0, \boldsymbol{ \ell } )$. Performing the above integrals for $a,b>0$ (for more details, see Appendix \ref{AppendixB}), we finally get:
\begin{align}
& \langle \rho_a\,\rho_b \rangle_\text{2loop}^{(2)}(\mathrm{i}\,\omega)
=- \frac{ 7\,e^2 \, m^{3-\frac{\varepsilon }{2}}\, | \omega | ^{1-\frac{\epsilon }{2}} 
}
{ 600\, \pi ^3 \,c}
 \left[\frac{1}{\varepsilon }-\frac{1}{2} 
\ln \left(\frac{m \,| \omega | }{\Lambda^2 }\right)\right]\,\delta_{ab}\,.
\end{align}

\subsubsection{Scaling of the Raman response at $T=0$}

In order to obtain any renormalized physical observable in the effective field theory formalism, we need to use the fact that $\frac{1} {\varepsilon^2}$ terms are canceled out by the corresponding counterterms of the renormalized action \cite{Peskin}. We also use the value ${m\,{e^*}^2}
/ \left( { \pi^2\, c} \right)= { 60\,\varepsilon} /{ 19 }$ at the NFL fixed point. Putting together all the terms, the final expression for $\langle \rho_a\,\rho_b \rangle(\mathrm{i}\,\omega)$ up to two-loop order takes the form:
\begin{align}
\langle \rho_a\,\rho_b \rangle(\mathrm{i}\,\omega)
& = \langle \rho_a\,\rho_b \rangle_\text{1loop}(\mathrm{i}\,\omega)
+ \langle \rho_a\,\rho_b \rangle_\text{2loop}^{(1)}(\mathrm{i}\,\omega) 
+\langle \rho_a\,\rho_b \rangle_\text{2loop}^{(2)}(\mathrm{i}\,\omega) 
+\langle \rho_a\,\rho_b \rangle_\text{counterterms} (\mathrm{i}\,\omega) 
\nn &
=  -\frac{   m^{2-\frac{\varepsilon }{2}}\, |\omega|^{1-\frac{\varepsilon }{2}}
\,\delta_{ab} }{10\, \pi }
-\frac{{e^*}^2  \,m^{3-\frac{\varepsilon }{2}}\, | \omega | ^{1-\frac{\epsilon }{2}} 
}
{ 150\, \pi ^3 \,c}
\ln \left(\frac{m \,| \omega | }{\Lambda^2 }\right) \delta_{ab}
+ \frac{ 7\,{e^*}^2 \, m^{3-\frac{\varepsilon }{2}}\, | \omega | ^{1-\frac{\epsilon }{2}} 
}    { 1200\, \pi ^3 \,c}
\ln \left(\frac{m \,| \omega | }{\Lambda^2 }\right) \delta_{ab}
\nn &
= -\frac{  m^{2-\frac{\varepsilon }{2}}\, |\omega|^{1-\frac{\varepsilon }{2}}
\,\delta_{ab} }{10\, \pi }
- 
\frac{ \varepsilon\, m^{2-\frac{\varepsilon }{2}} \,
| \omega | ^{1-\frac{\varepsilon }{2}} \ln \left(\frac{m\, | \omega | }{\Lambda^2 }\right)
\,\delta_{ab}}
{ 380 \,\pi }\nn
&=-\frac{   m^{2-\frac{\varepsilon }{2}}\, |\omega|^{1-\frac{\varepsilon }{2}}
\,\delta_{ab} }{10\, \pi }
\left[ 1 + 
\frac{ \varepsilon\,\ln \left(\frac{m\, | \omega | }{\Lambda^2 }\right)}
{ 38}
\right]\nn
& \simeq
-\frac{  m^{2-\frac{\varepsilon }{2}}\, 
|\omega|^{1-\frac{\varepsilon }{2}+\frac{ \varepsilon }
{ 38}}
\,\delta_{ab} }{10\, \pi }
\left(  \frac{m}{\Lambda^2}\right)^{ \frac{ \varepsilon }
{ 38 }}\,,
\end{align}
after re-exponentiating the correction term coming from the two-loop diagrams.
Therefore, the corrected Raman response scales as $|\omega|^{1-\varepsilon /2
+ \varepsilon / 38 }
\overset{\epsilon=1}  =
|\omega|^{1/2 + 1/38 } $, after including the leading-order corrections.

\subsection{Raman response at $T>0$}\label{Raman_finiteT}

In this subsection, we compute the Raman response at finite temperatures, using the memory matrix approach.
This approach is appropriate to compute finite temperature responses in systems where quasiparticles do not exist, and has been explained in great detail in our previous works (e.g., Refs.~\cite{Freire-AP_2014,Freire-AP_2017,Freire-AP_2020,ips_hermann1,ips-hermann2}). Just as in Refs. \cite{ips_hermann1,ips-hermann2}, we consider here the coupling to weak short-ranged disorder as the main mechanism for momentum relaxation in the LAB phase. Hence, we add the impurity term $H_{imp}= \int d^3\mathbf{x}\,W(\mathbf{x})
\,\psi^{\dagger}(t, \mathbf{x})\,\psi ( t, \mathbf{x})$ to the system.
We assume the disorder to have a Gaussian distribution, such that
$\left \langle\left \langle  W(\mathbf{x}) \right \rangle\right \rangle =0$ and $ 
\left \langle \left \langle W(\mathbf{x})\,W(\mathbf{x'}) \right \rangle\right \rangle  = W^2_0 \,\delta(\mathbf{x-x'})$,
where $W^2_0$ represents the average of the magnitude-squared of the random potential experienced by the fermions. We point out that, unlike the $T=0$ calculation (where we performed a systematic $\varepsilon$-expansion), we will now work directly in $d = 3$ in order to
minimize the technical complexity involved. In this formalism, we will then need to implement a
hard ultraviolet (UV) cutoff $\Lambda$ for the the momentum integrals, rather than using the dimensional regularization scheme (where the integrals are evaluated by sending the upper limit to $\infty$).

The $T>0$ Raman response can be conventionally defined as
\begin{align}
D_{\text{Raman}}(\omega,T)=\int_{0}^{\infty}dt \,e^{i\omega t}
\left \langle \left[\rho_a(t),\rho_b(0)
\right] \right \rangle.
\end{align}
This correlation function can be approximately written within the memory matrix formalism as
(cf. Refs.~\cite{Berg-PRB,wang2020low})
\begin{align}
\label{D_Raman}
D_{\text{Raman}}(\omega,T)\approx \chi_{\rho_a\rho_b}
\left(\frac{\mathrm{i}\,\omega}
{{M_{\rho_b\rho_b}(\omega})-\mathrm{i}\,\omega
\,\chi_{\rho_b\rho_b}}\right) \chi_{\rho_b \rho_a},
\end{align}
where $\chi_{\rho_a\rho_b}$ is the static thermodynamic susceptibility given by $\chi_{\rho_a\rho_b}(T)=\int_{0}^{\beta}d\tau \langle \rho_a(\tau) \,\rho_b(0) \rangle$, and $M_{\rho_a\rho_b}(\omega)$ is the memory matrix. The latter is defined by
\begin{align}\label{memory_matrix_definition}
M_{\rho_a\rho_b}(\omega,T)=
\int_{0}^{\beta}d\tau \left\langle \dot{\rho}_a(0)
\,\mathcal{Q}
\,\frac{\mathrm{i}}{\omega-\mathcal{Q}\,\text{L}\,\mathcal{Q}} 
\,\mathcal{Q} \,\dot{\rho}_b(\mathrm{i} \,\tau) \right\rangle,
\end{align}
where $\text{L}$ is the Liouville operator, and $\mathcal{Q}$ is an operator that projects out of a subspace of nearly-conserved operators.

\begin{figure}[t]
\includegraphics[width=0.45\textwidth]{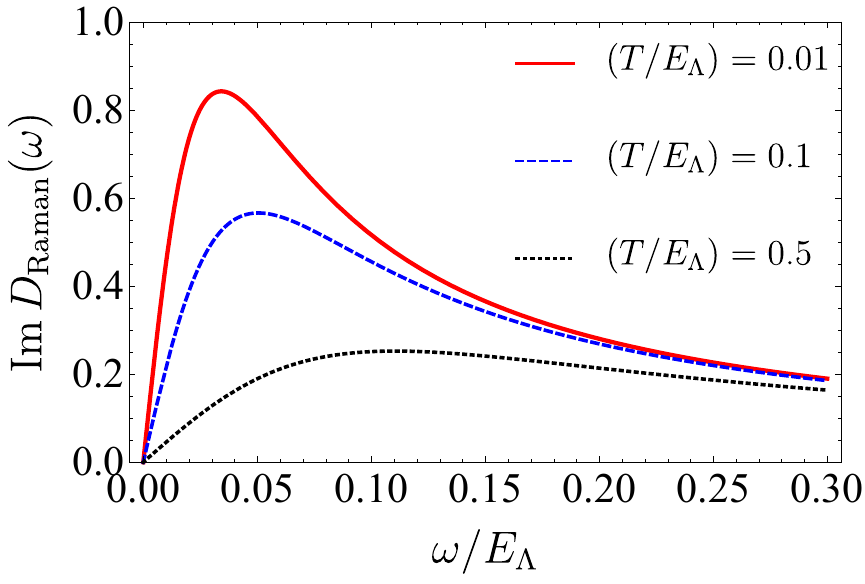}
\caption{\label{Raman_fig}
Plot of the Raman response as a function of frequency $\omega$ for three different values of the temperature $T$. We have used the parameter values of $\Lambda=10$, $m=1$, $m'=2$, $e^2/c=1$, and $W_0=1$. The variables $T$ and $\omega$ have been expressed in units of $E_{\Lambda} = \Lambda^2/(2\,m)$, which is a material-dependent ultraviolet energy cutoff, estimated by the energy scales upto which the dispersions of the conduction and valence bands scale quadratically with momentum (in the Luttinger semimetal compounds). 
} 
\end{figure}

We note that an important step in the memory matrix formalism is to identify a set of nearly-conserved operators in the system and, subsequently, project the dynamics onto these operators. As explained in Refs.~\cite{ips_hermann1,ips-hermann2}, the momentum operator $\mathbf{P}$ is a nearly-conserved operator in the system under consideration, that plays a key role in the computation of many transport coefficients in the system within the hydrodynamic regime. However, the momentum operator turns out to have no influence on the Raman response of the system, since it has no overlap with the operator $\rho_a$. Fortunately, we observe that $\rho_a$ is also a nearly-conserved operator,
\footnote{
When we calculate the equation of motion of $\rho_a$ due to the effects of weak disorder, we obtain
\begin{align}
\dot{\rho}_a=\mathrm{i}
\left [H + H_{imp},\rho_a \right ]
=\mathrm{i}
\int\frac{d^3 \mathbf{k}}{(2 \,\pi)^3} \int\frac{d^3 \mathbf{q}}{(2\pi)^3} \,
W(\mathbf{q})\left[\psi^\dagger({\mathbf{k+q}})\,\Gamma_a\, \psi ({\mathbf{k}})
-\psi^\dagger ({\mathbf{k}})\,\Gamma_a \,\psi ({\mathbf{k-q}})\right],
\end{align}
where $W(\mathbf{q})$ is the Fourier transform of the impurity coupling strength $W(\mathbf{x})$ defined in the text. Therefore, after performing the disorder averages, we can see that the operator $\rho_a$ is indeed a nearly-conserved operator, in the limit of weak disorder.}
and hence we can use it in place of $\mathbf{P}$. Consequently, to leading order, the memory matrix is given by $ M_{\rho_a\rho_b}(\omega,T)
\approx
{\text{Im}\,G^R_{\dot{\rho}_a\dot{\rho}_b}(\omega,T)}/{\omega}$ for small $\omega $-values,
where $G^R_{\dot{\rho}_a\dot{\rho}_b}(\omega,T)
= \langle \dot{\rho}_a(\omega) \dot{\rho}_b(-\omega)\rangle_0$ is the retarded correlation function for $\dot{\rho}_a= 
\mathrm{i} \left [H +H_{imp},\rho_a \right ]$. Here, $H$ represents the Hamiltonian of the system. The notation $\langle\cdots \rangle_0$ implies that, after the addition of the weak disorder term, the average in the grand-canonical ensemble, to leading order, is taken by employing the non-interacting part of the Hamiltonian. This gives the memory matrix as
\begin{align}
M_{\rho_a\rho_b}(\omega,T) \approx W_0^2
\int\frac{d^3 \mathbf{q}} {(2\,\pi)^3}\frac{\text{Im}\,
\Pi_{ab}^R(\mathbf{q},\omega)}{\omega}\,,
\end{align}
where $\Pi_{ab}^R(\mathbf{q},\omega)= \Pi(\mathbf{q},\mathrm{i}\,\omega
\rightarrow \omega + \mathrm{i} \,0^+)$, and
$\Pi_{ab}(\mathbf{q},\mathrm{i}\,\omega) = - T 
\sum \limits_{k_0}\int\frac{d^3 \mathbf{k}}{(2\pi)^3}
\text{Tr}[\Gamma_a\, G_0({k+q})\, \Gamma_b \,G_0({k})]$. Using Eq.~\eqref{D_Raman} and $\chi_{\rho_a\rho_b}\propto \delta_{ab}$, we finally obtain
\begin{align}
\text{Im}\,D_{\text{Raman}}(\omega,T)
=\frac{\omega \,M_{\rho_a \rho_a}
\,\chi^2_{\rho_a\rho_a}}
{\omega^2 \,\chi^2_{\rho_a\rho_a} + M^2_{\rho_a\rho_a}}
\equiv
\frac{\omega\,\widetilde{\Gamma}\,\chi_{\rho_a\rho_a}}
{\omega^2+\widetilde{\Gamma}^2}\,,
\end{align}
where $\widetilde{\Gamma}=M_{\rho_a\rho_a} \chi^{-1}_{\rho_a\rho_a}$. Since analytical expressions of $M_{\rho_a \rho_a}(T)$ and $\chi_{\rho_a \rho_a}(T)$ at finite temperatures cannot be obtained in closed-forms, we find their values using standard numerical integration techniques. We then fit the corresponding data in order to extract the temperature dependence of these quantities. This allows us to obtain $\text{Im}\,D_{\text{Raman}}(\omega,T)$, and a representative plot is shown in Fig.~\ref{Raman_fig}. From this figure, we observe that the Raman response at $T>0$ displays a quasi-elastic peak (QEP) at $\omega \approx \omega_{\text{max}}(T) =  M_{\rho_a\rho_a}(\omega=0,T)/\chi_{\rho_a\rho_a}(T)$, with the corresponding peak-height given by $ \chi_{\rho_a\rho_a}(T)/2$. To leading order, the static thermodynamic susceptibility is given by $\chi_{\rho_a\rho_a}(T)\sim a_1 + a_2 \,T^{1/2}+ a_3 /T$ (where $a_1$, $a_2 $, and $ a_3 $ are temperature independent constants). Furthermore, our numerical results show that $M_{\rho_a\rho_a}(\omega=0,T)$ is either a constant, or has an extremely weak $T$-dependence (i.e., not clearly observed within the numerical accuracy).

\section{Free energy at $T>0$}
\label{Free_energy}

In this section, we determine the scaling behavior of the
free energy at a finite temperature $T>0$, first from general scaling arguments, and
subsequently using the fixed point theory for the LAB phase, obtained using dimensional regularization \cite{moon-xu}.

\subsection{General arguments}

We derive the scaling forms for the case when hyperscaling property is obeyed.
In a system with spatial dimension $d$, dynamical critical
exponent $z$, the free energy $F$ has the scaling dimension
\begin{align}
[F] =d+z\,, 
\end{align}
and we expect the temperature-dependence to be
\begin{align}
F(T) \sim T^{d/z +  1}\,.
\label{eqfscale}
\end{align}
From this scaling form, we conclude that the specific heat should scale as
$C(T) \sim T^{d/z} $.
Since the current operator is given by $ J_\alpha(T) =\frac{\delta F}{\delta A_\alpha}$, where $A_\alpha$ is the vector potential with scaling dimension equal to unity, the Kubo formula gives the scaling dimension of the optical conductivity $\sigma (\omega)$ as
\begin{align}
[\sigma] &= 2\,[\mathbf J] - z- [\text{Volume in energy-momentum space}]\nonumber\\
&= 2\left( d+z-1\right)-z-d-z =d-2\,.
\end{align}
This implies the scaling form
\begin{align}
\sigma (\omega) \sim \omega^{(d-2)/z} \,.
\end{align}
We showed in Ref.~\cite{ips_hermann1} that this hyperscaling relation for $\sigma$ is violated, when we include the effects of Coulomb interactions at the LAB fixed point.

\vspace{0.3cm}

\subsection{Scaling of the free energy}

In this subsection, we compute the free energy density at a finite temperature $T>0$ to leading order in $\varepsilon $. The free energy density receives contributions from two parts, namely, the free fermions and the corrections due to the Coulomb interactions. Since the bosons $\varphi $ have no dynamics, there is no contribution from any free bosons, unlike other cases studied earlier by one of the authors \cite{ips-subir,ips-c2}.

The contribution of the free fermions is given by $\Delta F^{(0)} (T)=F^{(0)} (T) - F^{(0)} (0)$, such that
\begin{align}
\Delta F^{(0)} (T)
= - 2 \int \frac{ d^{4-\varepsilon}\mathbf k} {(2\,\pi)^{ 4-\varepsilon}}
\,\Bigg [ T \sum \limits_{p=\pm}\ln \left( 1 +  e^{- \frac{p\,|\mathbf k|^2 } 
{ 2\,m \, T} }\right) 
- \frac{ |\mathbf k|^2}{2\,m} \Bigg ]\,,
\end{align}
where we have subtracted the infinite contribution from the temperature-independent ground state energy, in order to make the expression finite. We have also included a factor of two to account for the doubly-degenerate bands. Rescaling $\mathbf k \rightarrow \mathbf k \,\sqrt{ 2\,m\, T }$, we get
\begin{align}
 \Delta F^{(0)} (T)
&= - \frac{ 2 \,\pi^{\frac{4-\varepsilon}{2}}
\left( 2\,m \, T \right)^{ 3-\frac{\varepsilon} {2}} 
} 
{ m\, \Gamma\left( \frac{4-\varepsilon}{2} \right)}
\int_0^\infty \frac{d |\mathbf k|\, |\mathbf k|^{3-\varepsilon}} 
{(2\,\pi)^{ 4-\varepsilon}}
\,\Bigg [ \sum \limits_{p=\pm}\ln \left( 1 +  e^{- p\,|\mathbf k|^2 } \right) 
- |\mathbf k|^2 \Bigg ] \nn
&= - \frac{ \pi^{\frac{4-\varepsilon}{2}}
\left( 2\,m \, T \right)^{ 3-\frac{\varepsilon} {2} } 
} 
{ m\, \Gamma\left( \frac{4-\varepsilon}{2} \right)}
\int_0^\infty \frac{du\, u^{ 1-\varepsilon/2}} 
{(2\,\pi)^{ 4-\varepsilon}}
\,\Bigg [ \ln \left( \frac{\left( 1 +  e^{u } \right) \left( 1 +  e^{-u } \right) 
}  {e^u}\right)  \Bigg ]
\nn & \simeq - \frac{  \eta(3)
\left(   m \, T \right)^{ 3-\frac{\varepsilon} {2} } } { \pi^2 \,m}
=  - \frac{ 3\, \zeta(3)
\left(   m \, T \right)^{ 3-\frac{\varepsilon} {2} } }
 { 4\, \pi^2 \,m} \,,
\end{align}
where
$ \eta(u)= \frac{\int_0^\infty dt \,t^{u-2} \,\ln(1+e^{-u})}
{\Gamma(u-1)}$ is the Dirichlet eta function, and we have set $\varepsilon= 0$ in the numerical prefactor. We have also used the identity $\eta (3)=\frac{3 \,\zeta (3)}{4}$, where $\zeta (u)$ is the Riemann zeta function.

\begin{figure}
\includegraphics[width=0.075\textwidth]{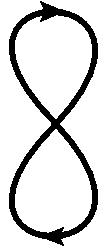}
\caption{The leading order correction to the free energy.
\label{fig:free_en}}
\end{figure}

The first order interaction correction to the free energy, as shown in Fig.~\ref{fig:free_en}, is given by
\begin{align}
F_{coul}
& =   \frac{e^2\,\Lambda^{\varepsilon}\,T^2}  {c }
 \sum_{\Omega_p, \,\omega_{p' } } 
 \int \frac{   d^{ 4-\varepsilon} \mathbf q \,
   d^{ 4-\varepsilon} \mathbf k  }   {  (2 \, \pi)^{ 8- 2\, \varepsilon } }
\frac{ \mathrm{Tr}\left  [ 
 G ( \omega_{p' } + \Omega_p, \mathbf k + \mathbf  q)  
 \, G( \omega_{p' }, \mathbf  k)\right ]} {\mathbf q^2}\,,
\end{align}
where $\Omega_p$ and $\omega_{p'}$ are bosonic and fermionic Matsubara frequencies, respectively. From this expression, we need to isolate the pole (in $\epsilon$) contributions. To lowest order in $\varepsilon$, these are obtained by evaluating
one frequency sum as an integral in the limit $T \rightarrow 0$, and the other one at finite temperature ($T>0$) (cf. Ref.~\cite{PhysRevB.92.165105}). To evaluate the frequency summations, we use the
following zeta-function regularization identities:
\begin{align}
T\sum \limits_{\Omega_p} \frac{1}{ |\Omega_p|^s}
=\frac{2\,T^{1-s} \,\zeta(s)}{(2\,\pi)^s} \,,\quad
T\sum \limits_{\omega_{p}} \frac{1}{ |\omega_p|^s}
=\frac{2\,T^{1-s} \,\zeta(s,1/2)}{(2\,\pi)^s}\,.
\end{align}

In case $\Omega_p$ becomes the continuous frequency for the zero temperature part,
we can rewrite the diagram as a fermionic loop with an insertion of the fermionic self-energy at $T = 0$, which is given by $\Sigma_1 (\mathbf k ) \equiv \Sigma_1 (\mathbf k,T=0 ) 
= -\frac{ m\,e^2  } 
{15\,\pi^2\,c} \left( \frac{\Lambda }{|\mathbf k |} \right)^\varepsilon  
\frac{ \mathbf{d}_{\mathbf k } \cdot \mathbf{\Gamma} }
{\varepsilon}$
(see Refs.~\cite{rahul-sid,*ips-rahul,*ips-rahul-errata}). 
We denote this as $F_{coul}^{(1)}$, which can be evaluated to
\begin{align}
F_{coul}^{(1)} &= - T\,\sum \limits_{\omega_{p'} } 
 \int \frac{   d^{ 4-\varepsilon} \mathbf k  }   {  (2 \, \pi)^{ 4- \varepsilon } }
 \mathrm{Tr}\left  [ \Sigma_1 ( \mathbf k )  
 \, G( \omega_{p' }, \mathbf  k)\right ]
= \frac{   T \,m\,e^2\,\Lambda^{\varepsilon}  } 
{ 15 \,\pi^2\,c\,\varepsilon } \sum_{\omega_{p' } }
 \int \frac{   d^{ 4-\varepsilon} \mathbf k  }   {  (2 \, \pi)^{ 4- \varepsilon } }
\frac{ \mathrm{Tr}\left  [ 
\left \lbrace \mathbf{d}_{\mathbf k } \cdot \mathbf{\Gamma} \right \rbrace 
  G( \omega_{p' }, \mathbf  k)\right ] }
{ |\mathbf k|^{\varepsilon} } \nn
& =  \frac{  4\, T \,m\,e^2\,\Lambda^{\varepsilon}  } 
{15\,\pi^2\,c\,\varepsilon } \sum_{\omega_{p' } }
 \int \frac{   d^{ 4-\varepsilon} \mathbf k  }   {  (2 \, \pi)^{ 4- \varepsilon } }
\frac{  \left | \mathbf{d}_{\mathbf k } \right |^2 }
{ |\mathbf k|^{\varepsilon} 
\left[  \omega_{p'}^2 +|\mathbf{d}(\mathbf{k})|^2   \right ]
} \nn
& = - \frac{e^2 \,\Lambda^{\varepsilon} \,m^{3-\varepsilon}  
\,\pi ^{\varepsilon -3} 
\csc \left(\frac{\pi  \varepsilon }{2}\right) }  { 30\, c\,\varepsilon}\,
T \sum_{\omega_{p' } } \frac{1}{|\omega_{p'}|^{\varepsilon-2} }\nn
&  \simeq
-\frac{e^2\,\Lambda^{\varepsilon} \,  \zeta (3) 
\left (m \, T \right )^{3-\varepsilon }}
{ 10 \,\pi ^4\, c\,\varepsilon}  \quad 
\left[\text{using } \zeta \left(\epsilon -2,\frac{1}{2}\right)
= \frac{-3 \, \epsilon \,\zeta '(-2)}{4} 
\text{ and }
\zeta '(-2) = -\frac{\zeta (3)}{4 \,\pi ^2}
\right ] 
\nonumber\\
&= 
-\frac{e^2   \,\zeta (3) 
\left( m\,T \right) ^{3-\frac{\varepsilon }{2}} }
{ 10 \, \pi ^4\, c \,\varepsilon }
+ \frac{e^2 \, \zeta (3) 
\left( m\,T \right) ^{3-\frac{\varepsilon }{2}} 
\ln \left(\frac{m \,T} {\Lambda^2 }\right)}
{ 20 \,\pi ^4 \,c}\,,
\end{align}
where we have set $\varepsilon= 0$ in the numerical prefactor.

The second case is when $\omega_{p'}$ becomes the continuous frequency. Then the contribution is
\begin{align}
& F_{coul}^{(2)} 
=   \frac{ 4\,e^2\,\Lambda^{\varepsilon} \,T}  {c }
 \sum_{\Omega_p} \int \frac{ d\omega_{p'} } {2\,\pi}
 \int \frac{   d^{ 4-\varepsilon} \mathbf q \,
   d^{ 4-\varepsilon} \mathbf k  }   {  (2 \, \pi)^{ 8- 2\, \varepsilon } }
\frac{ 
\left[-\left (  \omega_{p'} +  \Omega_{p}\right ) \omega_{p'}
+  \mathbf{d}_{\mathbf{k}+\mathbf q} \cdot \mathbf{d}_{\mathbf{k}} \right]} 
{\mathbf q^2
\left [ \left( \omega_{p'} + \Omega_p  \right )^2 
+|\mathbf{d}_{\mathbf{k}+\mathbf q}|^2\right ]
\left(  \omega_{p'} +|\mathbf{d}_{\mathbf{k}}|^2\right )} \nn
& =   \frac{ 4\,e^2\,\Lambda^{\varepsilon} \,T}  {c }
 \sum_{\Omega_p} 
 \int \frac{   d^{ 4-\varepsilon} \mathbf q \,
   d^{ 4-\varepsilon} \mathbf k  }   {  (2 \, \pi)^{ 8- 2\, \varepsilon } }\,
\frac{ \left[ -\mathbf{d}_{\mathbf{k}}
+
\frac{ 
d \left \lbrace \mathbf{k}\cdot \left( \mathbf k +\mathbf q \right)  \right \rbrace^2 
- \mathbf  k^2\left( \mathbf k +\mathbf q \right)^2 }  
{ 4\,m^2 \left( d-1\right)\, \mathbf{d}_{\mathbf{k}} } \right]}
{ {\mathbf q}^2 
\left[ (\mathbf{d}_{\mathbf{k}}+\mathbf{d}_{\mathbf{k}+\mathbf q})^2+\Omega ^2\right ]
}\,,
\end{align}
using $ \mathbf{d}_{\mathbf{k}} \cdot  \mathbf{d}_{\mathbf{p}} = 
 \frac{  d \left (\mathbf{k}\cdot\mathbf{p}  \right )^2 - \mathbf  k^2\,\mathbf  p^2 } 
 {4\,m^2 \left( d-1\right) } $.
Let $\theta$ be the angle between ${\mathbf{k}}$ and ${\mathbf{q}}$. Making a change of variables as $ |\mathbf{q}| = l\, |\mathbf{k}|$ (where $0< l < \infty $), and defining
$ \text{ang} = \sqrt{1+l^2+ 2\,l \cos \theta}$, we get
\begin{align}
 F_{coul}^{(2)} 
 & =   \frac{ 4\,e^2\,\Lambda^{\varepsilon} \,m^{3-\varepsilon }\,T}  
{c }
 \sum_{\Omega_p} \int 
\frac{   dl\, l^{ 3-\varepsilon}  \,d\theta } {(2 \, \pi)^{ 4- \varepsilon }} \,
\frac{ 
 \sin \theta \,l^{1-\varepsilon }  \left [ 2 \,l \left (\varepsilon -4 \right) \cos \theta 
 \left  ( l \cos \theta+2 \right )
+\left(\text{ang}^2-1\right) (5-\varepsilon )\right ]}
{
2^{ 8-\varepsilon}
\left( 3-\varepsilon\right) \pi ^{ 3-2 \varepsilon} 
\left(\text{ang}^2+1\right)^{4- \varepsilon}
\sin \left(\frac{\pi  \varepsilon }{2}\right)
| \Omega | ^{\varepsilon - 2} }\nn &
\simeq 
-\frac{ 19\,e^2\,\Lambda^{\varepsilon}\,  \zeta(3)
\left( m\,T \right) ^{3-\varepsilon }}  
{576 \,\pi ^4 \,c } + \mathcal{O} \left( \varepsilon \right) \,,
\end{align}
where we have set $\varepsilon= 0$ in the numerical prefactor. To this order of approximation, this part of the interaction correction to
the free energy does not contain a pole in $\varepsilon$, and hence $F_{coul}^{(2)} $ does
not contribute to the renormalization of the free energy.

Using $ {e^*}^2= \frac{ 60\,\pi^2 c\,\varepsilon} 
{19\,m}  $, the final expression for the free energy up to two-loop order takes the form
\begin{align}
& F^{(0)} (T ) - F^{(0)} (0 ) + F_{coul}^{(1)}
+F_\text{counterterms}^{(1)}
\nonumber\\
&=  - \frac{ 3\, \zeta(3)
\left(   m \, T \right)^{ 3-\frac{\varepsilon} {2} } }
 { 4\, \pi^2 \,m} 
+ \frac{ {e^*}^2 \, \zeta (3) 
\left( m\,T \right) ^{3-\frac{\varepsilon }{2}} 
\ln \left(\frac{m \,T} {\Lambda^2 }\right)}
{ 20 \,\pi ^4 \,c} 
\nn & 
=  - \frac{ 3\, \zeta(3)
\left(   m \, T \right)^{ 3-\frac{\varepsilon} {2} } }
 { 4\, \pi^2 \,m} 
\left [1 -  \frac{ 4 \,\varepsilon}
{ 19 } \,\ln \left(\frac{m \,T} {\Lambda^2 }\right)
\right]  \nn
& \simeq 
- \frac{ 3\, \zeta(3) 
\,  m ^{ 2-\frac{\varepsilon} {2} }\,
T^{3 -\frac{\varepsilon} {2} -  \frac{ 4\,\varepsilon}{ 19 }}
}
 { 4\, \pi^2 } 
\left( \frac {\Lambda^2 }{m } \right)^{\frac{ 4\,\varepsilon}{ 19}}\,,
\end{align}
after re-exponentiating the correction term coming from the two-loop diagrams and using Eq.~\eqref{eqe}. Note that the term $F_\text{counterterms}^{(1)} $ indicates that we have included the counterterms obtained from one-loop corrections in order to cancel out the singular $ \frac{1} {\varepsilon}$ terms in the renormalized action \cite{Peskin}.
Therefore, after including the leading order corrections,
the free energy scales as 
\begin{equation}
F(T)\sim  T^{3 -\frac{\varepsilon} {2} -  \frac{ 4\,\varepsilon}{ 19 }}
\overset{\epsilon=1}  =
T^{ \frac{5} {2} -  \frac{4}{ 19 }}
\,.
\label{eqEscale}
\end{equation}
This implies that the specific heat scales as
\begin{equation}
C(T)\sim  T^{ 2 -\frac{\varepsilon} {2} -  \frac{ 4\,\varepsilon}{ 19}}
\overset{\epsilon=1}  =
T^{ \frac{3} {2} -  \frac{4}{ 19 }}\,.
\label{eqcv}
\end{equation}
Comparing with Eq.~\eqref{eqfscale}, and taking into account the fixed point value $z^*=2-\frac{4\,\varepsilon} {19 }$ for the dynamical critical exponent,  we find that there is a hyperscaling violation, which is proportional to $\varepsilon$.
Finally, the entropy density $s$ is, by definition, the derivative of the free energy with respect to $T$. Hence, using Eq.~\eqref{eqEscale}, we infer that
\begin{equation}
s(T) \sim  T^{ 2 -\frac{\varepsilon} {2} -  \frac{ 4\,\varepsilon}{ 19 }}\overset{\epsilon=1}=T^{2-\frac{27}{38}}\,.
\label{eqs_scale}
\end{equation}
 
\section{Shear viscosity}
\label{shear_visc}

The momentum flux density tensor, also called the stress tensor, is given by
\begin{align}
\label{eqn:texpr}
T_{\mu\nu}=\sum_M \Big [\frac{\delta \mathcal{L}}{\delta (\partial_\mu \zeta_M)}\partial_\nu \zeta_M
-\partial_{\mu}
\Big \lbrace  \frac{\delta  \mathcal L}  {\delta \left(  \partial_\alpha \partial^\alpha \zeta_M \right) }
\Big \rbrace  \partial_\nu \zeta_M
\Big ]
-\delta_{\mu \nu} \, \mathcal{L} \,,
\end{align}
where $\mathcal L$ is the Lagrangian density, and $\zeta_M$ stands for all the quasiparticle fields in the theory.
Evaluating this explicitly for the Luttinger semimetal, we get
\begin{align}
T_{\mu\nu} (q_0,\mathbf q)
&=  \int\frac{ dk_0 \, d^d\mathbf{k}  }
 {(2\,\pi)^4}\,
\left( k_{\nu} + q_{\nu} /2 \right)\tilde{\psi} ^{\dagger}( k_0+q_0, \mathbf{k} +\mathbf q)
 \left [\nabla_{k_{\mu}}\mathbf{d_{k}}\cdot\mathbf{\Gamma} \right ]
\tilde{\psi} ( k_0, \mathbf{k}) \,.
\end{align}

The shear viscosity is the transport coefficient, which characterizes the relaxation of a transverse
momentum gradient back to local equilibrium.
Using the Kubo formula \cite{Taylor,zwerger}, the optical shear viscosity is given by
\begin{align}
 \eta (\omega)&=\lim_{\mathbf{q}\to 0} \frac{1} {\omega}\,
 \, \chi_{T_{xy} \, T_{xy}}(\omega,\mathbf{q})=\lim_{\mathbf{q}\to 0} \frac{1} {\omega}\,
 \, \chi_{T_{yz} \, T_{yz}}(\omega,\mathbf{q})
 =\lim_{\mathbf{q}\to 0} \frac{1} {\omega}\,\chi_{T_{zx} \, T_{zx}}(\omega,\mathbf{q}) \,, 
\end{align}
where
\begin{align}
 \chi_{T_{\alpha \beta } \, T_{ \alpha \beta}}(\omega,\mathbf{q})
 =  \langle T_{ \alpha \beta} \, T_{ \alpha \beta}\rangle (\omega, \mathbf q) 
\end{align}
is the autocorrelation function of the component $T_{\alpha \beta}$ of the stress tensor.

\begin{figure}[t]
	\centering
\includegraphics[width=0.3\textwidth]{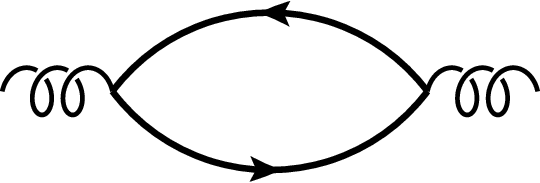} \label{fd1}
\caption{Feynman diagram for the contribution to the autocorrelation function
$\chi_{T_{\alpha \beta } \, T_{ \alpha \beta}}(\omega,\mathbf{q})$ at one-loop order.}
\label{fig1loop}
\end{figure}

\subsection{General arguments}

The spatial components of $T_{\alpha \beta }$ have the same scaling dimension as the Lagrangian density, and hence
\begin{align}
[T_{zx}]=d+z\,,
\end{align}
which immediately leads to
\begin{align}
[\eta] &= 2\,[T_{zx}] - z- [\text{Volume in energy-momentum space}]
= 2\left( d+z \right)-z-d-z =d \,.
\label{eqeta}
\end{align}
This shows that $\eta$ has the same scaling dimension as the entropy
density $s$, and hence the ratio $\eta/s$ is dimensionless, as expected.
If hyperscaling is not violated, Eq.~\eqref{eqeta} also implies the scaling form
\begin{align}
\eta (\omega) \sim \omega^{ d/z} 
\end{align}
for optical viscosity (i.e., for $ \omega \gg T $), and 
\begin{align}
\eta (T) \sim T^{ d/z} 
\end{align}
for dc viscosity (i.e., for $T\gg \omega $).

\subsection{Optical viscosity}
 In this subsection, we will find the expression of the optical viscosity at $T=0$, using the fixed point theory.

\subsubsection{One-loop contribution}

The autocorrelation function $\chi_{T_{\alpha \beta } \, T_{ \alpha \beta}}(\omega,\mathbf{0})$ at one-loop level is given by a simple fermionic 
loop with two insertions of the $T_{zx}$ operator, as shown in Fig.~\ref{fig1loop}. For the Luttinger semimetal, it evaluates to
\begin{align}
& \langle T_{zx} \, T_{zx} \rangle_\text{1loop}(\mathrm{i}\,\omega)
= -  \int \frac{dk_0}{2\,\pi} 
\int d ^{4-\epsilon} {\mathbf{k}} \,k_x^2 \,
\text{Tr} \left [ \left \lbrace 
\partial_{k_z}\mathbf{d}_{\mathbf k}\cdot \mathbf \Gamma\right \rbrace
G_0(k+q) \left \lbrace 
\partial_{k_z}\mathbf{d}_{\mathbf k}\cdot \mathbf \Gamma\right \rbrace G_0(k)\right ]\nn
&=  - \int \frac{dk_0}{2\,\pi} \int d ^{4-\epsilon} {\mathbf{k}}
 \,k_x^2 \,
\text{Tr} \left [ \left \lbrace 
\partial_{k_z}\mathbf{d}_{\mathbf k}\cdot \mathbf \Gamma\right \rbrace
\frac{ \mathrm{i}\, k_0 +\mathrm{i}\,\omega + \mathbf{d}_{\mathbf{k}} \cdot{\mathbf{\Gamma}}}
{-\left ( \mathrm{i}\, k_0 +\mathrm{i}\,\omega \right )^2 +|\mathbf{d}_{\mathbf{k}}|^2}\,
 \left \lbrace 
\partial_{k_z}\mathbf{d}_{\mathbf k}\cdot \mathbf  \Gamma\right \rbrace
\frac{ \mathrm{i}\, k_0  + \mathbf{d}_{\mathbf{k}} \cdot{\mathbf{\Gamma}}}
{-\left ( \mathrm{i}\, k_0   \right )^2 +|\mathbf{d}_{\mathbf{k}}|^2}
\right ]\nn
&= 4
  \int \frac{dk_0}{2\,\pi} \int d ^{4-\epsilon} {\mathbf{k}} \,k_x^2 \,
\frac{   \left \lbrace 
\partial_{k_z}\mathbf{d}_{\mathbf k}\right \rbrace^2\,( k_0 +\omega)\,k_0 
-  \frac{1}{2} \left \lbrace 
\partial_{k_z}  \mathbf{d}^2_{\mathbf k}\right \rbrace^ 2
+  \left \lbrace 
\partial_{k_z}\mathbf{d}_{\mathbf k}\right \rbrace^2 \mathbf{d}^2_{\mathbf k} }
{ \left [ -\left ( \mathrm{i}\, k_0 +\mathrm{i}\,\omega \right )^2 +|\mathbf{d}_{\mathbf{k}}|^2
\right ] 
\left [ -\left ( \mathrm{i}\, k_0   \right )^2 +|\mathbf{d}_{\mathbf{k}}|^2 \right] }\nn
& = - \frac{ m^{ 2-\frac{\varepsilon }{2}} \,
| \omega | ^{3-\frac{\varepsilon }{2}}}
{ 40 \,\pi   }\,,
\label{eq:tt_1Loop}
\end{align}
where $q=(\omega,\,\mathbf 0)$.
Consequently, at zeroth order, the optical viscosity $\eta(\omega)$ is proportional to $\omega^{ 2-\frac{\varepsilon }{2}}$ for $\omega\gg T$. For $d= 4-\varepsilon $, this scaling then obeys the hyperscaling $\omega^{d/z}$, as obtained from general arguments. 

\subsubsection{Two-loop contributions}

\begin{figure*}[t]
	\centering
	\subfigure[]{\includegraphics[width=0.3\textwidth]{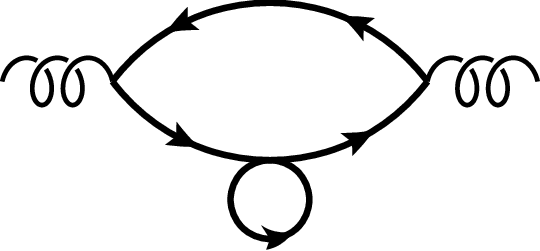} \label{fig2}}
	\subfigure[]{\includegraphics[width=0.3\textwidth]{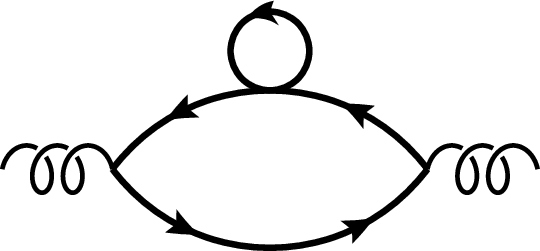} \label{fig3}}
	\subfigure[]{\includegraphics[width=0.3\textwidth]{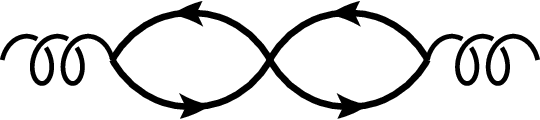} \label{fig4}}
	\caption{\label{fig2loop}Feynman diagrams for the contributions to the autocorrelation function $\chi_{T_{\alpha \beta } \, T_{ \alpha \beta}}(\omega,\mathbf{q})$ at two-loop order, with (a) and (b) representing the self-energy corrections, and (c) depicting the vertex correction.}
\end{figure*}

At two-loop order, we need to consider three Feynman diagrams, as shown in Figs.~\ref{fig2}, \ref{fig3}, and \ref{fig4}. The first two diagrams (Figs.~\ref{fig2} and \ref{fig3}) include the fermion self-energy corrections.
Evaluating them explicitly, we get
\begin{align}
& \langle T_{zx} \, T_{zx} \rangle_\text{2loop}^{(1)}(\mathrm{i}\,\omega) 
\nn & =
- 2  \int \frac{dk_0}{2\,\pi} 
\int \frac{d ^{4-\epsilon} {\mathbf{k}}}
{(2\,\pi)^{d}} \,k_x^2 \,
\text{Tr} \left [ \left \lbrace 
\partial_{k_z}\mathbf{d}_{\mathbf k}\cdot \mathbf \Gamma\right \rbrace
G_0(k+q) \,
\Sigma_1(k+q)
G_0(k+q)\left \lbrace 
\partial_{k_z}\mathbf{d}_{\mathbf k}\cdot \mathbf \Gamma\right \rbrace 
 \,G_0(k)
\right ] \nn
&=  
-\frac{e^2 \,  m^{2-\frac{\varepsilon }{2}} \,
| \omega | ^{3-\frac{\varepsilon }{2}}}
{ 675\, \pi ^3 \,c \,\varepsilon} 
\left( \frac{\Lambda^2}{m\,|\omega|} \right)^{\varepsilon/2} 
+ \frac{e^2 \,  m^{2-\frac{\varepsilon }{2}}
\, | \omega | ^{3-\frac{\varepsilon }{2}} 
\ln\left(\frac{m \,| \omega | }{\Lambda^2 }\right)}
{ 1350\, \pi ^3 \,c\, \varepsilon }\,.
 \end{align}
Note that we have included a factor of $2$, since the Figs.~\ref{fig2} and \ref{fig3} give equal contributions. The calculation of the loop integrals follows the same steps
as the computation of the corresponding two-loop current-current correlator in Ref.~\cite{ips_hermann1}. 

The diagram in Fig.~\ref{fig4} is the one with a vertex correction, and it evaluates to
\begin{align}
& 
\langle T_{zx} \, T_{zx} \rangle_\text{2loop}^{(2)}(\mathrm{i}\,\omega) 
\nn & =  \frac{e^2\,\Lambda^{\varepsilon}} {c}
\int \frac{dk_0\,d\ell_0}  {(2\,\pi)^2 }
\int \frac{d^{4-\epsilon} {\mathbf k}\, 
d^{4-\epsilon} {\boldsymbol{\ell}}
}
{(2\,\pi)^{2\,(4-\epsilon)}}
\,k_x \left(k_x+\ell_x \right)
\nn
& \hspace{ 3.2 cm}
\times\frac{  
\text{Tr} \left [ \left \lbrace 
\partial_{{k_z}}
\mathbf{d}_{\mathbf k}\cdot \mathbf \Gamma\right \rbrace
G_0(k_0+\omega,\mathbf k) \,
G_0( \ell_0 + \omega,\mathbf k+\boldsymbol \ell)
\left \lbrace 
\partial_{k_z+\ell_z}
\mathbf{d}_{\mathbf k + \boldsymbol{\ell}}\cdot \mathbf \Gamma\right \rbrace 
G_0(\ell_0,\mathbf k +\boldsymbol \ell)
\,G_0(k_0,\mathbf k)
\right ]} 
{\boldsymbol{\ell}^2}.
\end{align}
Again, the details of the evaluation of the loop integrals follow the same steps
as the computation of the corresponding two-loop current-current correlator in Ref.~\cite{ips_hermann1}, finally leading to 
\begin{align}
\langle T_{zx} \, T_{zx} \rangle_\text{2loop}^{(2)}(\mathrm{i}\,\omega) 
= 
-\frac{37 \,e^2\, m^{3-\frac{\varepsilon }{2}} \,
| \omega | ^{3-\frac{\varepsilon }{2}}}
{ 57600 \,\pi ^3 \,c \, \varepsilon }
\left( \frac{\Lambda^2 }{m\,|\omega|} \right)^{\varepsilon/2} 
+ 
\frac{37 \,e^2  \,m^{3-\frac{\varepsilon }{2}} 
\,| \omega | ^{3-\frac{\varepsilon }{2}} 
\ln \left (\frac{m \,| \omega |} {\Lambda^2 } \right )}
{ 115200 \,\pi ^3 \,c}\,.
\end{align}

\subsubsection{Scaling of optical viscosity at $T=0$}

Including the counterterms of the renormalized action \cite{Peskin}, using Eq.~\eqref{eqe} at the LAB fixed point, the final expression for the autocorrelator of the stress tensor $T_{zx}$ takes the form:
\begin{align}
\left \langle T_{zx} \,T_{zx} \right \rangle (\mathrm{i}\,\omega)
& = \left \langle T_{zx} \,T_{zx} \right \rangle _\text{1loop}(\mathrm{i}\,\omega)
+ \left \langle T_{zx} \,T_{zx} \right \rangle _\text{2loop}^{(1)}(\mathrm{i}\,\omega) 
+\left \langle T_{zx} \,T_{zx} \right \rangle _\text{2loop}^{(2)}(\mathrm{i}\,\omega) 
+\left \langle T_{zx} \,T_{zx} \right \rangle _\text{counterterms} (\mathrm{i}\,\omega) 
\nn &
=  - 
\frac{  m^{ 2-\frac{\varepsilon }{2}} \,
| \omega | ^{3-\frac{\varepsilon }{2}}}  { 40 \,\pi }
+ \frac{ {e^*}^2 \, m^{ 3-\frac{\varepsilon }{2}}
\, | \omega | ^{3-\frac{\varepsilon }{2}} 
\ln\left(\frac{m \,| \omega | }{\Lambda^2 }\right)}
{ 1350\, \pi ^3 \,c\, \varepsilon }
+ 
\frac{37 \,{e^*}^2\,  m^{3-\frac{\varepsilon }{2}} 
\,| \omega | ^{3-\frac{\varepsilon }{2}} 
\ln \left (\frac{m \,| \omega |} {\Lambda^2} \right )}
{ 115200 \,\pi ^3 \,c}\nn
& =  - \frac{  m^{ 2-\frac{\varepsilon }{2}} \,
| \omega | ^{3-\frac{\varepsilon }{2}}}  { 40 \,\pi }
\left[ 1- \frac{ 367\, \varepsilon }
{ 2736 } \, \ln \left (\frac{m \,| \omega |} {\Lambda^2} \right )
\right]\nn
& =  - \frac{  m^{ 2-\frac{\varepsilon }{2}} \,
| \omega | ^{3-\frac{\varepsilon }{2}
-\frac{ 367\, \varepsilon } {2736}}
}  
{ 40 \,\pi }
 \, \left ( \frac{\Lambda^2} {m}\right )
 ^{\frac{ 367\, \varepsilon }
{ 2736 }} \,,
\end{align}
up to two-loop order. As usual, we have re-exponentiated the correction term coming from the two-loop diagrams. Therefore, the corrected optical viscosity in the clean limit scales as 
\begin{equation}
\eta(\omega)\sim \omega^{2-\frac{\varepsilon }{2}
-\frac{ 367\, \varepsilon } {2736}}\,,
\end{equation}
after including the leading order corrections due to the presence of the Coulomb interactions. Comparing with Eq.~\eqref{eqeta}, we find that there is indeed a small hyperscaling violation proportional to $\varepsilon$. 
 
\subsection{DC Viscosity}

We now focus on studying the shear viscosity $\eta(\omega=0,T)$ at $T>0$. Following the same strategy explained previouly in the context of the Raman response at $T>0$, we will use the memory matrix method, and work directly in $d=3$ spatial dimensions. Moreover, we will also assume here the presence of short-ranged weak disorder (that breaks translational invariance) as the source of momentum relaxation. In the memory matrix formalism, the shear viscosity is given by
\begin{align}
\label{eta}
\eta(\omega,T)= \chi_{T_{ \alpha \beta} T_{ \alpha \beta}}(T)
\left[ \frac{1}
{M_{T_{ \alpha \beta} T_{ \alpha \beta}}(T)-\mathrm{i}\,\omega
\,\chi_{T_{\alpha \beta} T_{\alpha \beta}}(T)}\right ] 
\chi_{T_{\alpha \beta} T_{\alpha \beta}}(T)
\,,
\end{align}
where $\chi_{T_{\alpha \beta } \, T_{ \alpha \beta}}(T)
=\int_{0}^{\beta}d\tau 
\left \langle T_{\alpha \beta}(\tau,\mathbf{q}=0) \,T_{\alpha \beta}(0,\mathbf{q}=0) \right  \rangle$ 
is the static susceptibility of the component $T_{\alpha \beta}$ of the stress tensor, and $M_{T_{ \alpha \beta} T_{ \alpha \beta}}(T)$ is the corresponding memory matrix element [this is defined in a way similar to the one defined in Eq.~\eqref{memory_matrix_definition}]. We want to emphasize that the stress tensor is also a nearly-conserved operator for the Luttinger semimetal, because the equation of motion of $T_{ \alpha \beta}$ due to the effects of weak disorder is 
\begin{align}
\dot{T}_{\alpha \beta}
& =
\mathrm{i} \left [H+H_{imp},T_{ \alpha \beta} \right ] \nn
\Rightarrow
\dot{T}_{\alpha \beta}
& =
\mathrm{i} \,
\int\frac{d^3 \mathbf{k}}{(2\pi)^3} \int\frac{d^3 \mathbf{k'}}{(2\pi)^3} \,k_{\beta} \,\tilde W(\mathbf{k'})
\left[\psi^\dagger ({\mathbf{k+k'}})\,(\partial_{k_{\alpha}}\mathbf{d_{k}\cdot\Gamma})\, \psi ({\mathbf{k}})
-\psi^\dagger ({\mathbf{k}})\,(\partial_{k_{\alpha}}\mathbf{d_{k}\cdot\Gamma}) \,\psi ({\mathbf{k-k'}})\right],
\end{align}
where $\tilde W(\mathbf{k'})$ is the Fourier transform of the impurity coupling strength $W(\mathbf{x})$ (defined in Sec.~\ref{Raman_finiteT}).

The susceptibility $\chi_{T_{zx} \, T_{ zx}}^{\text{1loop}}(T)$ at one-loop level is given by a simple fermionic 
loop with two insertions of the $T_{zx}$ operator, as shown in Fig.~\ref{fig1loop}, and hence is captured by
\begin{align}
&\chi_{T_{zx} \, T_{zx}}^{\text{1loop}}(T)
= - T\sum_{k_0}
\int \frac{d ^{3}{\mathbf{k}}}{(2\pi)^3} \,k_x^2 \,
\text{Tr} \left [ \left \lbrace 
\partial_{k_z}\mathbf{d}_{\mathbf k}\cdot \mathbf \Gamma\right \rbrace
G_0(k) \left \lbrace 
\partial_{k_z}\mathbf{d}_{\mathbf k}\cdot \mathbf \Gamma\right \rbrace G_0(k)\right ]\nn
&= 4\,
  T\sum_{k_0}\int \frac{d ^{3}{\mathbf{k}}}{(2\pi)^3} \,k_x^2 \,
\frac{   \left \lbrace 
\partial_{k_z}\mathbf{d}_{\mathbf k}\right \rbrace^2\,k_0^2 
-  \frac{1}{2} \left \lbrace 
\partial_{k_z}  \mathbf{d}^2_{\mathbf k}\right \rbrace^ 2
+  \left \lbrace 
\partial_{k_z}\mathbf{d}_{\mathbf k}\right \rbrace^2 \mathbf{d}^2_{\mathbf k} }
{ \left [ -\left ( \mathrm{i}\, k_0  \right )^2 +|\mathbf{d}_{\mathbf{k}}|^2
\right ]^2 }\,.
\label{eq:tt_1Loop}
\end{align}

At two-loop order, we need to consider three Feynman diagrams, as shown in Figs.~\ref{fig2}, \ref{fig3}, and \ref{fig4}. We will denote the sum of these contributions by $\chi_{T_{zx} \, T_{zx}}^{\text{2loop}}(T)=\chi^{(2,1)}_{T_{zx} \, T_{zx}}(T)
+\chi^{(2,2)}_{T_{zx} \, T_{zx}}(T)$, such that $\chi^{(2,1)}_{T_{zx} \, T_{zx}}(T)$
is the sum of first two diagrams [cf. Figs.~\ref{fig2} and \ref{fig3}] which include the fermion self-energy corrections.

We find that
\begin{align}
&\chi^{(2,1)}_{T_{zx} \, T_{zx}}(T)= - 2\,T\sum_{k_0}
\int \frac{d ^{3}{\mathbf{k}}}{(2\pi)^3} \,k_x^2 \,
\text{Tr} \left [ \left \lbrace 
\partial_{k_z}\mathbf{d}_{\mathbf k}\cdot \mathbf \Gamma\right \rbrace
G_0(k+q) \,
\Sigma_T(k+q)
G_0(k+q)\left \lbrace 
\partial_{k_z}\mathbf{d}_{\mathbf k}\cdot \mathbf \Gamma\right \rbrace 
 \,G_0(k)
\right ]\, , 
\end{align}
where the temperature-dependent contribution of the self-energy is
$\Sigma_T(\mathbf{k})
=
-\frac{e^2 \Lambda}{15\pi^2 c}\frac{(\mathbf{d}_{\mathbf k}\cdot \mathbf{\Gamma})}{T}
$. Note that we have included a factor of $2$ in the expression above, since the Figs.~\ref{fig2} and \ref{fig3} give equal contributions. This calculation involves steps similar to those involved in the
computation of the corresponding two-loop diagram for current-momentum susceptibility in Ref.~\cite{ips_hermann1}.

The diagram in Fig.~\ref{fig4} is the one with a vertex correction, and it evaluates to
\begin{align}
&\chi^{(2,2)}_{T_{zx} \, T_{zx}}(T)
 \nn & =  \frac{e^2\, T^2} {c}
\sum_{k_0, \,\ell_0} 
\int \frac{d^{3} {\mathbf k}\, 
d^{3} {\boldsymbol{\ell}}
}
{(2\,\pi)^{6}}
\,k_x \left(k_x+\ell_x \right)
\frac{  
\text{Tr} \left [ \left \lbrace 
\partial_{{k_z}}
\mathbf{d}_{\mathbf k}\cdot \mathbf \Gamma\right \rbrace
G_0(k_0,\mathbf k) 
G_0( \ell_0 ,\mathbf k+\boldsymbol \ell)
\left \lbrace 
\partial_{k_z+\ell_z}
\mathbf{d}_{\mathbf k + \boldsymbol{\ell}}\cdot \mathbf \Gamma\right \rbrace 
G_0(\ell_0,\mathbf k +\boldsymbol \ell)
\,G_0(k_0,\mathbf k)
\right ]} 
{\boldsymbol{\ell}^2}.
\end{align}
Again, the details of the above calculation follow steps similar to those in Ref.~\cite{ips_hermann1}. We would like to emphasize that although an analytical expression of $\chi_{T_{zx} \, T_{zx}}^{\text{2loop}}(T)$
cannot be obtained in a closed-form, its value is estimated by using standard numerical integration methods. We use the numerical data into a fitting function in order to extract the temperature dependence, which gives us $\chi_{T_{zx} \, T_{zx}}(T)=\mathcal{A}+\mathcal{B}\,T^{5/2}+\mathcal{C}/T$, where $\mathcal{A}$, $\mathcal{B}$, and $\mathcal{C}$ are temperature-independent constants. In fact, $\mathcal{A}$ and $\mathcal{C}$ are non-universal, since they depend on the hard UV momentum cutoff $\Lambda$. At low temperatures, we get $\chi_{T_{zx} \, T_{zx}}(T)\approx\mathcal{A}+\mathcal{C}/T$.

At leading order, the memory matrix element is given by
\begin{align}
& M_{T_{zx} T_{zx}}(T)\nn
&=-W_0^2 
\lim_{\omega\rightarrow 0}\frac{\text{Im}
\bigg [
\int \frac{d^3 {\mathbf k} \, d^3 {\mathbf q}}{(2\pi)^6}
\,k_x^2\, T\sum \limits _{k_0}\text{Tr}\left[(\partial_{k_z+q_z}\mathbf{d}_{\mathbf{k+q}} \cdot{\mathbf{\Gamma}})\,G_0(\omega+k_0,\mathbf{k+q})\,(\partial_{k_z}\mathbf{d}_{\mathbf{k}} \cdot{\mathbf{\Gamma}})\,
G_0(k_0,\mathbf{k})\right]  \bigg ]
\Bigg|_{\mathrm{i} \,\omega\rightarrow\omega+\mathrm{i} \,\delta}}
{\omega}\, .
\end{align}
After computing the trace, we perform the analytical continuation, and the resulting integral is evaluated numerically. This yields $M_{T_{zx} T_{zx}}(T)\sim \mathcal{A'}+\mathcal{B'}/T$, with $\mathcal{A'}$ and $\mathcal{B'}$ being (non-universal) temperature-independent constants that depend only on the UV cutoff $\Lambda$. At low temperatures, we obtain $M_{T_{zx} T_{zx}}(T)\approx \mathcal{B'}/T$.

Using Eq. \eqref{eta}, in the $\omega\rightarrow 0$ limit, we get
$\eta(T)=\chi^2_{T_{zx} \, T_{zx}}(T)/M_{T_{zx} \, T_{zx}}(T)$. Therefore, at low temperatures,
\begin{align}
\label{eqetaT}
\eta(T)\sim T^{\lambda}, \text{ where }0<\lambda< 1 \,.
\end{align}

\subsection{Scaling of the ratio $\eta /s$ at $T>0$}

\label{secetabys}

Eqs.~\eqref{eqetaT} and \eqref{eqs_scale} give us the ratio
\begin{align}
\eta/s \sim T^{\lambda-\frac{49}{38}}
\end{align}
in $d=3$ spatial dimensions. The scaling implies that $\eta/s$ diverges at low temperatures, instead of saturating the universal bound. This divergent behavior is similar to what was found in NFL metals near the Ising-nematic critical point \cite{patel2}, where the quantum critical phase has a critical Fermi surface, and the divergence is a consequence of the hyperscaling violation due to the presence of this Fermi surface \cite{ips-subir}. We also note that this singular dependence of $\eta/s$ on $T$ is at variance with the results found by earlier works in Refs.~\cite{dumitrescu,Herbut-PRB}, where the authors used a quantum Boltzmann equation approach.

\section{Conclusion}
\label{secsum} 

In this paper, we have performed a two-loop calculation to determine the Raman scattering response at intermediate frequencies in the LAB phase. We have found that at frequencies larger than the temperature, the Raman response exhibits a power-law behavior, which can be verified experimentally. At low frequencies, the Raman response displays instead a clear quasi-elastic peak. In the second part, we have computed the ratio of the shear viscosity and the entropy density $\eta/s$. We have demonstrated that in the physically relevant case (i.e., $N_f=1$), this ratio diverges at low temperatures, instead of saturating the Kovtun-Son-Starinets universal bound. This divergent behavior has some similarities to what was found in the quantum critical phase near an Ising-nematic critical point \cite{patel2}. Consequently, it can be traced to the violation of the hyperscaling property that emerges in the LAB phase. 

From our results here, we can find the scaling of the thermal diffusivity $D_{th} =
{\kappa} / {C}$, where $\kappa$ is the thermal conductivity at zero current. Since the scalings of the specific heat and the thermal conductivity are described by $C \sim  
 T^{ 2 -\frac{\varepsilon} {2} -  \frac{ 4\,\varepsilon}{ 19 }}$ [cf. Eq.~\eqref{eqcv}] and $\kappa\sim u\,T^{-1}$ \cite{ips-hermann2} (where $u$ is a parametrically small prefactor), respectively, we get
 $ D_{th} \sim
u\,T^{ -3 +\frac{\varepsilon} {2} + \frac{ 4\,\varepsilon}{ 19 }} $.

\section*{Acknowledgments}

We thank Xiaoyu Wang for participating in the initial stages of the project. H.F. acknowledges funding from CNPq under grant numbers 310710/2018-9 and 311428/2021-5.

\bibliography{biblio_rev}

\appendix

\section{$d_a$-function algebra and other useful identities}
\label{angular}

We derive a set of useful relations \cite{lukas-herbut,igor16} for the vector functions $\mathbf{d}(\mathbf{k})$ (whose components $d_a(\mathbf{k})$ are the $l=2$ spherical harmonics in $d$ spatial dimensions) and the generalized real $d\times d$ Gell-Mann matrices $\Lambda_a$ ($a = 1,2,\cdots, N$) in $d$-dimensions.
The  matrices $\Lambda_a$ are  symmetric,  traceless, and orthogonal, satisfying
\begin{align}
& \text{Tr}[  \Lambda^a \,  \Lambda^b] =2\,\delta_{ab} \,,
\quad \sum_{a=1}^N \left ( \Lambda^a \right )_{ij} \left ( \Lambda^a_{l j'}\right )
=\delta_{i l} \, \delta_{j j'}+ \delta_{i j'}\, \delta_{jl}
- \frac{2} {d}\,\delta_{ i j }\, \delta_{l j'}\,.
\end{align}
As a result, the index $a$ (or $b$) runs from $1$ to $N = \frac{\left (d-1\right)\left (d+2\right)}{2}$.
We define the components of $\mathbf{d}(\mathbf k)$ by
\begin{align}
d_a(\mathbf{k}) =\sqrt { \frac{d}{2 \left(d-1 \right ) }}
\sum \limits_{i,j=1}^{d}
\frac{ k_i \left ( \Lambda^a\right)_{ij}  k _j } {2\,m}\,.
\end{align}
This gives 
\begin{align}
& \sum \limits _{a=1}^N 
 d_a(\mathbf{k}) \, d_a(\mathbf{p}) = 
 \frac{  d 
 \left (\mathbf{k}\cdot\mathbf{p}  \right )^2 - \mathbf  k^2\,\mathbf  p^2 } 
 {4\,m^2 \left( d-1\right) } \,.
 \label{eqdprod3}
\end{align}

The trace of the six gamma matrices, used for the loop-calculations, is given by
\begin{align}
\label{eq6trace}
\text{Tr}
\left(\Gamma_a\, \Gamma_b\, \Gamma_c\, \Gamma_d\, \Gamma_e\, \Gamma_f \right) 
= \,&4\,
\Big [\delta_{ab} \left (\delta_{cd}\,\delta_{ef}-\delta_{ce}\,\delta_{df}+\delta_{cf}\,\delta_{de} 
\right )
-\delta_{ac}\left (\delta_{bd}\,\delta_{ef}-\delta_{be}\,\delta_{df}+\delta_{bf}\,\delta_{de} 
\right )
 +
\delta_{ad}\left (\delta_{bc}\,\delta_{ef}-\delta_{be}\,\delta_{cf}+\delta_{bf}\,\delta_{ce} \right)
\nonumber\\
& \quad 
-\delta_{ae}\left (\delta_{bc}\,\delta_{df}-\delta_{bd}\,\delta_{cf}+\delta_{bf}\,\delta_{cd} \right)
+\delta_{af}\left (\delta_{bc}\,\delta_{de}-\delta_{bd}\,\delta_{ce}+\delta_{be}\,\delta_{cd}
\right )\Big] \,.
\end{align}

\section{Details of the two-loop calculations for the $T=0$ Raman response}
\label{AppendixB}

In this appendix, we provide the details of the two-loop calculations for the $T=0$ Raman response.

As explained in the main text, the first two-loop-order correction involves inserting the one-loop fermion self-energy ($\Sigma_1$) corrections into the correlator, which takes the form:
\begin{align}
 \left \langle \rho_a\,\rho_b \right \rangle_\text{2loop}^{(1)}(\mathrm{i}\,\omega)
&= -    \int \frac{dk_0}{2\,\pi} 
\int \frac{d^{4-\epsilon} {\mathbf k}}{(2\,\pi)^{4-\epsilon}}
\text{Tr} \left [ \Gamma_a\,G_0(k+q) \,\Sigma_1(\mathbf k)
\,G_0(k+q)\, \Gamma_b\, \,G_0(k)
\right ]\nn
&\quad -    \int \frac{dk_0}{2\,\pi} 
\int \frac{d^{4-\epsilon} {\mathbf k}}{(2\,\pi)^{4-\epsilon}}
\text{Tr} \left [ \Gamma_a\,G_0(k+q) \,\Gamma_b\,
G_0(k)\,\Sigma_1(\mathbf k)  \,G_0(k)
\right ]\,,
\end{align}
where $\Sigma_1 (\boldsymbol \ell) 
= -\frac{ m\,e^2  } 
{15\,\pi^2\,c} \left( \frac{\Lambda }{|\boldsymbol \ell|} \right)^\varepsilon  
\frac{ \mathbf{d}(\boldsymbol \ell) \cdot \mathbf{\Gamma} }
{\varepsilon}$ (see Ref.~\cite{rahul-sid,ips-rahul,*ips-rahul-errata}).
This gives us
\begin{align}
 \left \langle \rho_a\,\rho_b \right \rangle_\text{2loop}^{(1)}(\mathrm{i}\,\omega)
&= \frac{  m\,e^2 } 
{15\,\pi^2\,c\,\varepsilon }   
 \int \frac{dk_0}{2\,\pi} 
\int \frac{d^{4-\epsilon} {\mathbf k}}{(2\,\pi)^{4-\epsilon}} 
\left( \frac{\Lambda}{|\mathbf k|} \right)^\varepsilon
\frac{term_1
}
{
 \left\lbrace -\mathcal{B}^2 +|\mathbf{d}(\mathbf{k})|^2 \right \rbrace
\left \lbrace -\mathcal{A}^2 +|\mathbf{d}(\mathbf{k})|^2 \right \rbrace^2
}\nn
& \quad +
\frac{ m\,e^2 } 
{15\,\pi^2\,c\,\varepsilon }   
 \int \frac{dk_0} {2\,\pi} 
\int \frac{d^{4-\epsilon} {\mathbf k}}{(2\,\pi)^{4-\epsilon}} 
\left( \frac{\Lambda}{|\mathbf k|} \right)^\varepsilon
\frac{term_2
}
{\left\lbrace - \mathcal{B}^2 +|\mathbf{d}(\mathbf{k})|^2 \right \rbrace^2
\left \lbrace - \mathcal{A}^2 +|\mathbf{d}(\mathbf{k})|^2 \right \rbrace
}
\,,
\end{align}
where
\begin{align}
\mathcal{A} = \mathrm{i}\, k_0 +\mathrm{i}\,\omega - \frac{ \mathbf k^2}{2\,m'}\,,\quad
\mathcal{B} = \mathrm{i}\, k_0  - \frac{ \mathbf k^2}{2\,m'}\,,
\end{align}
\begin{align}
term_1 &=
\text{Tr} \left [\Gamma_a
\left \lbrace
\mathcal{A}  + \mathbf{d}(\mathbf{k}) \cdot{\mathbf{\Gamma}}
\right \rbrace 
\left \lbrace\mathbf{d}(\mathbf k)\cdot \mathbf{\Gamma} \right \rbrace
\left \lbrace
\mathcal{A} + \mathbf{d}(\mathbf{k}) \cdot{\mathbf{\Gamma}}
\right \rbrace 
 \Gamma_b
\left \lbrace
\mathcal{B} + \mathbf{d}(\mathbf{k}) \cdot{\mathbf{\Gamma}}
\right \rbrace  
\right],\nn
term_2 &=
\text{Tr} \left [\Gamma_a 
\left \lbrace
\mathcal{A} + \mathbf{d}(\mathbf{k}) \cdot{\mathbf{\Gamma}}
\right \rbrace 
 \Gamma_b
\left \lbrace
\mathcal{B} + \mathbf{d}(\mathbf{k}) \cdot{\mathbf{\Gamma}}
\right \rbrace  
\left \lbrace\mathbf{d}(\mathbf k)\cdot \mathbf{\Gamma} \right \rbrace
\left \lbrace
\mathcal{B}  + \mathbf{d}(\mathbf{k}) \cdot{\mathbf{\Gamma}}
\right \rbrace  
\right].
\end{align}

First, let us evaluate $term_1$ for $a=b=0$ as follows:
\begin{align}
& term_1 \nn &=
\text{Tr} \left [
\left \lbrace
\mathcal{A} + \mathbf{d}(\mathbf{k}) \cdot{\mathbf{\Gamma}}
\right \rbrace 
\left \lbrace\mathbf{d}(\mathbf k)\cdot \mathbf{\Gamma} \right \rbrace
\left \lbrace
\mathcal{A}  + \mathbf{d}(\mathbf{k}) \cdot{\mathbf{\Gamma}}
\right \rbrace 
\left \lbrace
\mathcal{B} + \mathbf{d}(\mathbf{k}) \cdot{\mathbf{\Gamma}}
\right \rbrace  \right]\nn
& =\mathcal{A} \,\text{Tr} \left [
\left \lbrace\mathbf{d}(\mathbf k)\cdot \mathbf{\Gamma} \right \rbrace
\left \lbrace
\mathcal{A}  + \mathbf{d}(\mathbf{k}) \cdot{\mathbf{\Gamma}}
\right \rbrace 
\left \lbrace
\mathcal{B} + \mathbf{d}(\mathbf{k}) \cdot{\mathbf{\Gamma}}
\right \rbrace  \right]
+ \text{Tr} \left [
\left \lbrace \mathbf{d}(\mathbf{k}) \cdot{\mathbf{\Gamma}}
\right \rbrace 
\left \lbrace\mathbf{d}(\mathbf k)\cdot \mathbf{\Gamma} \right \rbrace
\left \lbrace
\mathcal{A}  + \mathbf{d}(\mathbf{k}) \cdot{\mathbf{\Gamma}}
\right \rbrace 
\left \lbrace
\mathcal{B} + \mathbf{d}(\mathbf{k}) \cdot{\mathbf{\Gamma}}
\right \rbrace  \right]\nn
& = \left( \mathcal{A}^2 +2\, \mathcal{A} \,\mathcal{B} \right) \text{Tr} \left [
\left \lbrace\mathbf{d}(\mathbf k)\cdot \mathbf{\Gamma} \right \rbrace
\left \lbrace
 \mathbf{d}(\mathbf{k}) \cdot{\mathbf{\Gamma}}
\right \rbrace  \right]
+ \text{Tr} \left [
\left \lbrace \mathbf{d}(\mathbf{k}) \cdot{\mathbf{\Gamma}}
\right \rbrace 
\left \lbrace\mathbf{d}(\mathbf k)\cdot \mathbf{\Gamma} \right \rbrace
\left \lbrace
 \mathbf{d}(\mathbf{k}) \cdot{\mathbf{\Gamma}}
\right \rbrace 
\left \lbrace
 \mathbf{d}(\mathbf{k}) \cdot{\mathbf{\Gamma}}
\right \rbrace  \right]\nn
& = \left( \mathcal{A}^2 +2\, \mathcal{A} \,\mathcal{B} \right) \frac{\mathbf k^4}{m^2}
+ \frac{\mathbf k^8}{4\,m^4}\,.
\end{align}
Similarly, we get:
\begin{align}
term_2  = 
\left( \mathcal{B}^2 +2\, \mathcal{A} \,\mathcal{B} \right) \frac{\mathbf k^4}{m^2}
+ \frac{\mathbf k^8}{4\,m^4}\,.
\end{align}
Evaluating the integrals, we obtain:
\begin{align}
 \left \langle \rho_0\,\rho_0 \right \rangle_\text{2loop}^{(1)}(\mathrm{i}\,\omega)
&=   0\,.
\end{align}
Let us now evaluate $term_1$ for $a, b>0$ as follows:
\begin{align}
& term_1 \nn &=
\text{Tr} \left [\Gamma_a
\left \lbrace
\mathcal{A} + \mathbf{d}(\mathbf{k}) \cdot{\mathbf{\Gamma}}
\right \rbrace 
\left \lbrace\mathbf{d}(\mathbf k)\cdot \mathbf{\Gamma} \right \rbrace
\left \lbrace
\mathcal{A}  + \mathbf{d}(\mathbf{k}) \cdot{\mathbf{\Gamma}}
\right \rbrace 
 \Gamma_b
\left \lbrace
\mathcal{B} + \mathbf{d}(\mathbf{k}) \cdot{\mathbf{\Gamma}}
\right \rbrace  \right]\nn
&=
\mathcal{A}\,\text{Tr} \left [\Gamma_a 
\left \lbrace\mathbf{d}(\mathbf k)\cdot \mathbf{\Gamma} \right \rbrace
\left \lbrace
\mathcal{A}  + \mathbf{d}(\mathbf{k}) \cdot{\mathbf{\Gamma}}
\right \rbrace 
 \Gamma_b
\left \lbrace
\mathcal{B} + \mathbf{d}(\mathbf{k}) \cdot{\mathbf{\Gamma}}
\right \rbrace  \right]
+ 
\text{Tr} \left [\Gamma_a 
\left \lbrace \mathbf{d}(\mathbf{k}) \cdot{\mathbf{\Gamma}} \right \rbrace
\left \lbrace\mathbf{d}(\mathbf k)\cdot \mathbf{\Gamma} \right \rbrace
\left \lbrace
\mathcal{A}  + \mathbf{d}(\mathbf{k}) \cdot{\mathbf{\Gamma}}
\right \rbrace 
 \Gamma_b
\left \lbrace
\mathcal{B} + \mathbf{d}(\mathbf{k}) \cdot{\mathbf{\Gamma}}
\right \rbrace  \right] \nn
&=
2\,\mathcal{A}\,\mathcal{B}\, \text{Tr} \left [\Gamma_a 
\left \lbrace\mathbf{d}(\mathbf k)\cdot \mathbf{\Gamma} \right \rbrace
\left \lbrace
 \mathbf{d}(\mathbf{k}) \cdot{\mathbf{\Gamma}}
\right \rbrace 
 \Gamma_b  \right]
+
\mathcal{A}^2\,\text{Tr} \left [\Gamma_a 
\left \lbrace\mathbf{d}(\mathbf k)\cdot \mathbf{\Gamma} \right \rbrace
 \Gamma_b
\left \lbrace
 \mathbf{d}(\mathbf{k}) \cdot{\mathbf{\Gamma}}
\right \rbrace  \right]
 \nn & \quad 
+
\text{Tr} \left [\Gamma_a 
\left \lbrace \mathbf{d}(\mathbf{k}) \cdot{\mathbf{\Gamma}} \right \rbrace
\left \lbrace\mathbf{d}(\mathbf k)\cdot \mathbf{\Gamma} \right \rbrace
\left \lbrace
\mathbf{d}(\mathbf{k}) \cdot{\mathbf{\Gamma}}
\right \rbrace 
 \Gamma_b
\left \lbrace \mathbf{d}(\mathbf{k}) \cdot{\mathbf{\Gamma}}
\right \rbrace  \right]\nn
&=
2\,\mathcal{A}\,\mathcal{B}
\left | \mathbf{d}(\mathbf{k}) \right |^2 \text{Tr} \left [\Gamma_a\,\Gamma_b  \right]
+
\left(\mathcal{A}^2 +\left | \mathbf{d}(\mathbf{k}) \right |^2\right)
\text{Tr} \left [\Gamma_a 
\left \lbrace\mathbf{d}(\mathbf k)\cdot \mathbf{\Gamma} \right \rbrace
 \Gamma_b
\left \lbrace
 \mathbf{d}(\mathbf{k}) \cdot{\mathbf{\Gamma}}
\right \rbrace  \right]\nn
&=
8\,\mathcal{A}\,\mathcal{B}
\left | \mathbf{d}(\mathbf{k}) \right |^2 \delta_{ab}
+
4 \left(\mathcal{A}^2 +\left | \mathbf{d}(\mathbf{k}) \right |^2\right)
 \left [
2\,{d}_a(\mathbf k)\,{d}_b(\mathbf{k}) 
- \left | \mathbf{d}(\mathbf{k}) \right |^2 \delta_{ab} \right]
\nn &=
4 \left[ 2\,\mathcal{A}\,\mathcal{B}
-  \left(\mathcal{A}^2 +\left | \mathbf{d}(\mathbf{k}) \right |^2\right)
\right] \left | \mathbf{d}(\mathbf{k}) \right |^2 \delta_{ab} 
+
8 \left(\mathcal{A}^2 +\left | \mathbf{d}(\mathbf{k}) \right |^2\right)
 {d}_a(\mathbf k)\,{d}_b(\mathbf{k}) \,.
\end{align}
Similarly, for $a,b>0$, $term_2 $ evaluates to:
\begin{align}
& term_2 \nn & =
\text{Tr} \left [\Gamma_a 
\left \lbrace
\mathcal{A}  + \mathbf{d}(\mathbf{k}) \cdot{\mathbf{\Gamma}}
\right \rbrace 
 \Gamma_b
\left \lbrace
\mathcal{B}  + \mathbf{d}(\mathbf{k}) \cdot{\mathbf{\Gamma}}
\right \rbrace  
\left \lbrace\mathbf{d}(\mathbf k)\cdot \mathbf{\Gamma} \right \rbrace
\left \lbrace
\mathcal{B}  + \mathbf{d}(\mathbf{k}) \cdot{\mathbf{\Gamma}}
\right \rbrace  \right] \nn
& = \mathcal{B}\,\text{Tr} \left [\Gamma_a 
\left \lbrace
\mathcal{A}  + \mathbf{d}(\mathbf{k}) \cdot{\mathbf{\Gamma}}
\right \rbrace 
 \Gamma_b
\left \lbrace
\mathcal{B}  + \mathbf{d}(\mathbf{k}) \cdot{\mathbf{\Gamma}}
\right \rbrace  
\left \lbrace\mathbf{d}(\mathbf k)\cdot \mathbf{\Gamma} \right \rbrace \right]
+
\text{Tr} \left [\Gamma_a 
\left \lbrace
\mathcal{A}  + \mathbf{d}(\mathbf{k}) \cdot{\mathbf{\Gamma}}
\right \rbrace 
 \Gamma_b
\left \lbrace
\mathcal{B}  + \mathbf{d}(\mathbf{k}) \cdot{\mathbf{\Gamma}}
\right \rbrace  
\left \lbrace\mathbf{d}(\mathbf k)\cdot \mathbf{\Gamma} \right \rbrace
\left \lbrace \mathbf{d}(\mathbf{k}) \cdot{\mathbf{\Gamma}}
\right \rbrace  \right]\nn
& =\left( \mathcal{B}^2+ \left |\mathbf{d}(\mathbf k)\right |^2\right)
\text{Tr}  \left [\Gamma_a 
\left \lbrace \mathbf{d}(\mathbf{k}) \cdot{\mathbf{\Gamma}}
\right \rbrace 
 \Gamma_b   
\left \lbrace\mathbf{d}(\mathbf k)\cdot \mathbf{\Gamma} \right \rbrace \right]
+
2\,\mathcal{A} \,\mathcal{B}\,\left |\mathbf{d}(\mathbf k)\right |^2
\text{Tr} \left [\Gamma_a \, \Gamma_b \right]\nn
& =4
\left( \mathcal{B}^2+ \left |\mathbf{d}(\mathbf k)\right |^2\right)
 \left [ 2\,{d}_a(\mathbf{k})\,{d}_b(\mathbf{k})
-\left |\mathbf{d}(\mathbf k)\right |^2
\delta_{ab} \right]
+
8\,\mathcal{A} \,\mathcal{B}\,\left |\mathbf{d}(\mathbf k)\right |^2
\delta_{ab}\nn
&=
4 \left[ 2\,\mathcal{A}\,\mathcal{B}
-  \left(\mathcal{B}^2 +\left | \mathbf{d}(\mathbf{k}) \right |^2\right)
\right] \left | \mathbf{d}(\mathbf{k}) \right |^2 \delta_{ab} 
+
8 \left(\mathcal{B}^2 +\left | \mathbf{d}(\mathbf{k}) \right |^2\right)
 {d}_a(\mathbf k)\,{d}_b(\mathbf{k}) \,.
\end{align}
Evaluating the integrals, we finally obtain:
\begin{align}
 \left \langle \rho_a\,\rho_b \right \rangle_\text{2loop}^{(1)}(\mathrm{i}\,\omega)
&=   
\frac{e^2 \, m^{3-\frac{\varepsilon }{2}}\, | \omega | ^{1-\frac{\epsilon }{2}} 
}
{ 75\, \pi ^3 \,c}
 \left[\frac{1}{\varepsilon }-\frac{1}{2} 
\ln \left(\frac{m \,| \omega | }{\Lambda^2 }\right)\right]\,\delta_{ab}\,.
\end{align}

The second two-loop-order correction are the vertex-like corrections, which can be expressed as
\begin{align}
 & \left \langle \rho_a\,\rho_b \right \rangle_\text{2loop}^{(2)}(\mathrm{i}\,\omega)
 =  \frac{e^2\,\Lambda^{\varepsilon}} {c}
  \int \frac{dk_0\,d\ell_0}  {(2\,\pi)^2 }
\int \frac{d^{4-\epsilon} {\mathbf k}\, d^{4-\epsilon} {\boldsymbol{\ell}}}
{(2\,\pi)^{2\,(4-\epsilon)}
}
\text{Tr} \left [  \Gamma_a\,
G_0(k+q) \,\frac{1} {\boldsymbol{\ell}^2}\, G_0(k+q+\ell)\,
 \Gamma_b\, G_0(k+\ell)\,G_0(k) \right ]\nn
& = \frac{e^2\,\Lambda^{\varepsilon}} {c}
\int \frac{dk_0\,d\ell_0}  {(2\,\pi)^2 }
\int \frac{d^{4-\epsilon} {\mathbf k}\, d^{4-\epsilon} {\boldsymbol{\ell}}}
{(2\,\pi)^{2\,(4-\epsilon)}}
\text{Tr} \left [  \Gamma_a\,
G_0(k_0+\omega,\mathbf k) \,\frac{1} {\boldsymbol{\ell}^2}\, 
G_0( \ell_0 + \omega,\mathbf k+\boldsymbol \ell)\,
 \Gamma_b \,G_0(\ell_0,\mathbf k +\boldsymbol \ell)
\,G_0(k_0,\mathbf k)
\right ],
\end{align}
with $q=(\omega,\mathbf 0)$ and $\ell =(\ell_0, \boldsymbol{ \ell } )$.
After some convenient regrouping of the terms in the integrand,
the expression to be evaluated takes the form:
\begin{align}
& \langle \rho_a\,\rho_b \rangle_\text{2loop}^{(2)}(\mathrm{i}\,\omega)
\nn & =  \frac{ e^2\,\Lambda^{\varepsilon}} {c}
\int \frac{d^{4-\epsilon} {\mathbf k}\, d^{4-\epsilon} {\boldsymbol{\ell}}}
{(2\,\pi)^{2\,(4-\epsilon)}
}
\text{Tr} \left [ 
\frac{\int \frac{d \ell_0}  {2\,\pi }\,G_0(\ell_0-\omega,\boldsymbol \ell)
\,\Gamma_a\,G_0(\ell_0 ,\boldsymbol \ell) \,
\int \frac{dk_0}  {2\,\pi }\,
G_0(k_0+ \omega, \mathbf k)\,\Gamma_b\,
G_0( k_0, \mathbf k)}
{\left( \mathbf k +  \boldsymbol{\ell} \right )^2}
\right ].
\end{align}
It is clear from this expression that $\langle \rho_a\,\rho_b \rangle_\text{2loop}^{(2)}(\mathrm{i}\,\omega) \propto \delta_{ab}$. Hence, we will consider $\langle \rho_a\,\rho_a \rangle_\text{2loop}^{(2)}(\mathrm{i}\,\omega)$ (no sum over $a$).

\begin{enumerate}

\item $\langle \rho_a\,\rho_a \rangle_\text{2loop}^{(2)}(\mathrm{i}\,\omega)= 0$.

\item $\langle \rho_a\,\rho_0 \rangle_\text{2loop}^{(2)}(\mathrm{i}\,\omega)= 0$.

\item Next, let us consider $a,b>0$.
Let us evaluate the $k_0$-integral, such that
\begin{align}
& \int \frac{d k_0}  {2\,\pi }\,
G_0(k_0+ \omega,  \mathbf k)\,\Gamma_b\,
G_0(k_0, \mathbf k)
= \int \frac{d k_0}  {2\,\pi }\,
\frac{ \left \lbrace
\mathcal{A}  + \mathbf{d}( \mathbf k) \cdot{\mathbf{\Gamma}}
\right \rbrace \Gamma_b \left \lbrace
\mathcal{B}  + \mathbf{d}( \mathbf k) \cdot{\mathbf{\Gamma}}
\right \rbrace
}
{ \left( -\mathcal{A}^2 +  |\mathbf{d}( \mathbf k)|^2\right)
\left( -\mathcal{B}^2 +  |\mathbf{d}( \mathbf k)|^2\right)
}\nn
& =
\int \frac{d\ell_0}  {2\,\pi }\,
\frac{ 
\mathcal{A} \,\mathcal{B} \,\Gamma_b
+\mathcal{A} \,\Gamma_b
\left \lbrace\mathbf{d}( \mathbf k) \cdot{\mathbf{\Gamma}} \right \rbrace
+ \mathcal{B} \left \lbrace\mathbf{d}( \mathbf k) \cdot{\mathbf{\Gamma}} \right \rbrace \Gamma_b
+ \left \lbrace\mathbf{d}( \mathbf k) \cdot{\mathbf{\Gamma}} \right \rbrace
 \Gamma_b \left \lbrace\mathbf{d}( \mathbf k) \cdot{\mathbf{\Gamma}} \right \rbrace
}
{ \left( -\mathcal{A}^2 +  |\mathbf{d}( \mathbf k)|^2\right)
\left( -\mathcal{B}^2 +  |\mathbf{d}( \mathbf k)|^2\right)
}\nn
& = m^3\,
\frac{ 
-\frac{ \mathbf k^4} {2 \,m^2} \,\Gamma_b
-i\,\omega \,\Gamma_b
\left \lbrace\mathbf{d}( \mathbf k) \cdot{\mathbf{\Gamma}} \right \rbrace
+ i\,\omega \left \lbrace\mathbf{d}( \mathbf k) \cdot{\mathbf{\Gamma}} \right \rbrace \Gamma_b
+ 2 \left \lbrace\mathbf{d}( \mathbf k) \cdot{\mathbf{\Gamma}} \right \rbrace
 \Gamma_b \left \lbrace\mathbf{d}( \mathbf k) \cdot{\mathbf{\Gamma}} \right \rbrace
}
{ \mathbf k^2 \left( \mathbf k^4 + m^2 \,\omega ^2\right)
}\,.
\end{align}
Plugging this in, we get:
\begin{align}
& \langle \rho_a\,\rho_a \rangle_\text{2loop}^{(2)}(\mathrm{i}\,\omega)
 =  \frac{ e^2\,\Lambda^{\varepsilon}\,m^6} {c}
\int \frac{d^{4-\epsilon}{\mathbf k}\, d^{4-\epsilon} {\boldsymbol{\ell}}}
{(2\,\pi)^{2\,(4-\epsilon)}} 
\frac{ term_3}
{\left( \mathbf k +  \boldsymbol{\ell} \right )^2
\boldsymbol \ell^2 \, \mathbf k^2 
\left( \boldsymbol \ell^4 + m^2 \,\omega ^2\right)
 \left( \mathbf k^4 + m^2 \,\omega ^2\right)
}\,,
\end{align}
where
\begin{align}
& term_3 \nn &= \text{Tr} \Bigg[\,
\left \lbrace 
-\frac{ \boldsymbol \ell^4\,\Gamma_a} {2 \,m^2} 
+i\,\omega \,\Gamma_a
\left ( \mathbf{d}( \boldsymbol \ell) \cdot{\mathbf{\Gamma}} \right )
- i\,\omega \left (\mathbf{d}( \boldsymbol \ell) \cdot{\mathbf{\Gamma}} \right ) \Gamma_a
+ 2 \left (\mathbf{d}( \boldsymbol \ell) \cdot{\mathbf{\Gamma}} \right )
 \Gamma_a \left (\mathbf{d}(\boldsymbol \ell) \cdot{\mathbf{\Gamma}} \right )
\right \rbrace
\nn & \hspace{1 cm} 
\left \lbrace 
-\frac{ \mathbf k^4\,\Gamma_a} {2 \,m^2} 
-i\,\omega \,\Gamma_a
\left ( \mathbf{d}( \mathbf k) \cdot{\mathbf{\Gamma}} \right )
+ i\,\omega \left (\mathbf{d}( \mathbf k) \cdot{\mathbf{\Gamma}} \right ) \Gamma_a
+ 2 \left (\mathbf{d}( \mathbf k) \cdot{\mathbf{\Gamma}} \right )
 \Gamma_a \left (\mathbf{d}( \mathbf k) \cdot{\mathbf{\Gamma}} \right )
\right \rbrace \,\Bigg]\nn
& =
\text{Tr} \Bigg[\,
\frac{ \boldsymbol \ell^4\,\mathbf k^4} {4 \,m^4} 
-\frac{ \boldsymbol \ell^4\,
\Gamma_a\left (\mathbf{d}( \mathbf k) \cdot{\mathbf{\Gamma}} \right )
 \Gamma_a \left (\mathbf{d}(\mathbf k) \cdot{\mathbf{\Gamma}} \right )} 
{m^2}
- \frac{ \mathbf k^4\,
\Gamma_a\left (\mathbf{d}( \boldsymbol \ell) \cdot{\mathbf{\Gamma}} \right )
 \Gamma_a \left (\mathbf{d}(\boldsymbol \ell) \cdot{\mathbf{\Gamma}} \right )} 
{m^2}
+2\,\omega^2 \,\Gamma_a
\left ( \mathbf{d}( \boldsymbol \ell) \cdot{\mathbf{\Gamma}} \right ) 
\Gamma_a
\left ( \mathbf{d}( \mathbf k) \cdot{\mathbf{\Gamma}} \right )
\nn & \hspace{1 cm}
-2\,\omega^2\,\mathbf{d}( \boldsymbol \ell) \cdot \mathbf{d}( \mathbf k)
\,\Bigg] + \tilde term_3 \,,\nn
& = 4\, \Bigg[\,
\frac{ \boldsymbol \ell^4\,\mathbf k^4} {4 \,m^4} 
-\boldsymbol \ell^4\, \frac{ 
2\,{d}^2_a( \mathbf k) -|\mathbf{d}( \mathbf k)|^2 } 
{m^2}
- \mathbf k^4\,\frac{ 
2\,{d}^2_a( \boldsymbol \ell) -|\mathbf{d}( \boldsymbol \ell)|^2 
} 
{m^2}
+ 4\,\omega^2 \left\lbrace
{d}_a( \mathbf k)\, {d}_a( \boldsymbol \ell)
- \mathbf {d}( \mathbf k)\cdot \mathbf {d}( \boldsymbol \ell)
\right \rbrace \,\Bigg] + \tilde term_3  \,.
\end{align}
where
\begin{align}
& \tilde term_3
= 4 \,\text{Tr} \Bigg[\,  \left (\mathbf{d}( \boldsymbol \ell) \cdot{\mathbf{\Gamma}} \right )
 \Gamma_a \left (\mathbf{d}(\boldsymbol \ell) \cdot{\mathbf{\Gamma}} \right )
\left (\mathbf{d}( \mathbf k) \cdot{\mathbf{\Gamma}} \right )
 \Gamma_a \left (\mathbf{d}( \mathbf k) \cdot{\mathbf{\Gamma}} \right )
\,\Bigg] 
\nn &
= 16\left[
-\frac{ d^2_{a}(\boldsymbol \ell) \,\mathbf k^4
+ d^2_{a}(\mathbf k) \,\boldsymbol \ell^4} 
{ 2\,m^2 }
+ \frac{\mathbf k^4\, \boldsymbol \ell^4} {16\,m^4}
+ 4\,d_{a}(\boldsymbol \ell) \, d_{a}(\mathbf k)
\left \lbrace \mathbf d(\boldsymbol \ell) \cdot 
\mathbf d(\mathbf k) \right \rbrace
\right].
\end{align}

Hence, using the symmetry of the integral under the exchange $\mathbf k \leftrightarrow \boldsymbol \ell $, we can rewrite the integrand as
\begin{align}
& term_3 \nn & =
\frac{ \boldsymbol \ell^4\,\mathbf k^4} {m^4} 
-2\,\boldsymbol \ell^4\, \frac{ 
8\,{d}^2_a( \mathbf k) - \frac{ {\mathbf k}^4} {m^2} } 
{m^2}
+ 16\,\omega^2 \left\lbrace
{d}_a( \mathbf k)\, {d}_a( \boldsymbol \ell)
- \mathbf {d}( \mathbf k)\cdot \mathbf {d}( \boldsymbol \ell)
\right \rbrace 
- \frac{ 16\,d^2_{a}(\mathbf k) \, \boldsymbol \ell^4} { m^2}
+  \frac{ \boldsymbol \ell^4\, \mathbf k^4 } { m^4}
+ 64\,d_{a}(\boldsymbol \ell) \, d_{a}(\mathbf k)
\left \lbrace \mathbf d(\boldsymbol \ell) \cdot 
\mathbf d(\mathbf k) \right \rbrace \nn
& = \frac{ 2\, \boldsymbol \ell^4\,\mathbf k^4} {m^4} 
-2\,\boldsymbol \ell^4\, \frac{ 
16\,{d}^2_a( \mathbf k) - \frac{ {\mathbf k}^4} {m^2} } 
{m^2}
+ 16\,\omega^2 \left\lbrace
{d}_a( \mathbf k)\, {d}_a( \boldsymbol \ell)
- \mathbf {d}( \mathbf k)\cdot \mathbf {d}( \boldsymbol \ell)
\right \rbrace 
+ 64\,d_{a}(\boldsymbol \ell) \, d_{a}(\mathbf k)
\left \lbrace \mathbf d(\boldsymbol \ell) \cdot 
\mathbf d(\mathbf k) \right \rbrace .
\end{align}

\end{enumerate}

Finally, performing the integrals, we get:
\begin{align}
& \langle \rho_a\,\rho_b \rangle_\text{2loop}^{(2)}(\mathrm{i}\,\omega)
=- \frac{ 7\,e^2 \,m^{3-\frac{\varepsilon }{2}}\, | \omega | ^{1-\frac{\epsilon }{2}} 
}
{ 600\, \pi ^3 \,c}
 \left[\frac{1}{\varepsilon }-\frac{1}{2} 
\ln \left(\frac{m \,| \omega | }{\Lambda^2 }\right)\right]\,\delta_{ab}\,.
\end{align}

\end{document}